\documentclass[11pt]{article}
\usepackage{amsmath,amssymb,amsfonts,epsfig}

\newcommand{\wickn}[1]{\protect{:\!#1\!:}}
\newcommand{\wickq}[1]{\protect{
\begin{picture}(2,10)
\put(1,0){\circle{1}}
\put(1,3){\circle{1}}
\put(1,6){\circle{1}}
\end{picture}
\,#1\,
\begin{picture}(3,10)
\put(1,0){\circle{1}}
\put(1,3){\circle{1}}
\put(1,6){\circle{1}}
\end{picture}
}}

\newcommand{\mathscr}{\mathcal}

\newtheorem{prop}{Proposition}

\newtheorem{thm}[prop]{Theorem}

\newcommand{\proof}[1]{\vspace{1.5ex}{\small 
{\bf Proof:} #1 \hfill$\blacksquare$}\vspace{1.5ex}}

\newcommand{\example}[1]{\vspace{1.5ex}
{\bf Example:} #1 \vspace{1.5ex}}

\newcommand{\tsf}[2]{\protect{{\textstyle{\frac{ #1}{#2}}}}}

\newcommand{\ts}[1]{\protect{{\textstyle{ #1}}}}

\newcommand{\beqa}{\begin{eqnarray*}}
\newcommand{\eeqa}{\end{eqnarray*}}

\newcommand{\beqan}{\begin{eqnarray}}
\newcommand{\eeqan}{\end{eqnarray}}

\newcommand{\invF}[1]{\big(#1\big)\check{\phantom{l}}}

\newcommand{\cphi}{\varphi}          % the field on CST
\newcommand{\qphi}{\phi}             % the field on QST

\title{Field Theory on Noncommutative Spacetimes:
\\Quasiplanar Wick Products}

\author{D. Bahns$^*$, 
S. Doplicher$^\ddag$, 
K. Fredenhagen$^*$, G. Piacitelli$^\ddag$
\\
\\
$

\begin{array}{l}
\!\!\phantom{ }^*
\mbox{{\footnotesize II. Institut f\"ur Theoretische Physik, Universit\"at
Hamburg, Luruper Chaussee 149, }}
\\ \phantom{^*}\mbox{{\footnotesize D - 22761 Hamburg, Germany --
 {\tt bahns@mail.desy.de, fredenha@mail.desy.de}}}
\\\!\!\phantom{ }^\ddag\mbox{{\footnotesize Dipartimento di Matematica
, Universit\`a di Roma ``La Sapienza'',
P.le Aldo Moro 2, }}
\\\phantom{^\ddag}\mbox{{\footnotesize 00185 Roma, Italy --{
\tt dopliche@mat.uniroma1.it, piacitel@mat.uniroma1.it}}}
\end{array}
$}

\date{}

\begin{document}
\maketitle

\begin{abstract} 

\noindent We give a definition of admissible counterterms appropriate for
massive quantum field theories on the noncommutative Minkowski space, based 
on a
suitable notion of locality. We then define 
products of fields of arbitrary order, the so-called quasiplanar Wick
products,   by subtracting only such admissible counterterms. We derive the
analogue of Wick's theorem  and  comment on the consequences of using
quasiplanar Wick products in the perturbative expansion. 

\end{abstract}

\section{Introduction}

\noindent Interest in quantum field theories with nonlocal interactions has
reemerged recently in the context of the analysis of quantum field theory on
noncommutative spacetimes. Such spacetimes are studied for various reasons, one
of them based on the observation that Heisenberg's uncertainty principle along
with classical gravity suggests that the localization of an event in spacetime
with an arbitrarily high precision should be impossible. Based on this
argument, a noncommutative spacetime  (called the noncommutative Minkowski
space or quantum spacetime) was
introduced in~\cite{dfr}. Here, the ordinary coordinates are replaced by
noncommuting ``quantum coordinates'', i.e. selfadjoint operators $q^\mu$,
$\mu=0,\dots,3$, with 
\[ 
[q^\mu,q^\nu]=iQ^{\mu\nu}\ ,
\] 
subject to certain ``quantum conditions'', 
\[
[q^\rho, Q^{\mu\nu} ]=0\,,\qquad Q_{\mu\nu} Q^{\mu\nu}\,=\,0\,,\qquad
\left(\,\tsf 1 2 \, Q_{\mu\nu}\,Q_{\rho\sigma}\, \epsilon^{\,\mu\nu\rho\sigma}\,\right)^2 
\,=\,16\,\lambda_P^8 \, I 
\]
where $\lambda_P$ is the Planck length, such that for every state $\omega$ in
the domain of $[q^\mu,q^\nu]$ the following relations hold among the
uncertainties  $\Delta (q^\mu) = \sqrt{\omega ((q^\mu)^2) - \omega (q^\mu)^2}$
:
\begin{eqnarray*} 
\Delta q^0 \cdot
\left (\Delta q^1 + \Delta q^2 +\Delta q^3\right)  &\geq& 
\lambda_P^2
\\ 
\Delta q^1\cdot\Delta q^2 + \Delta q^1\cdot\Delta
q^3  + \Delta q^2 \cdot\Delta q^3&\geq& \lambda_P^2\ .
\end{eqnarray*} 
As shown in~\cite{dfr}, the regular realizations of the quantum conditions, i.e.
those satisfying
\[
e^{i\alpha q}e^{i\beta q}=e^{i(\alpha +\beta) q}e^{-\frac i 2 \alpha Q \beta }
\, , \quad \alpha, \beta \in \mathbb R^4\, , \; 
\alpha q= \alpha_\mu q^\mu\,,\,\alpha Q \beta=\alpha_{\mu} Q^{\mu\nu} \beta_\nu\ ,
\]
are in one-to-one correspondence with the non-degenerate representations of a
$C^*$-algebra which is isomorphic to the algebra 
$\mathcal E=C_0(\Sigma,\mathcal K)$, where
$\mathcal K$ is the algebra of 
compact operators on a fixed separable infinite dimensional Hilbert space and 
$\Sigma$ is the 
joint spectrum  of the operators $Q^{\mu\nu}$. This spectrum, being fixed in a
Poincar\'e-invariant way  by the quantum conditions, is homeomorphic to two
copies of the tangent bundle of the 2-sphere,  the 
noncompact manifold $TS^2\times \{1,-1\}$. The commutators $Q^{\mu\nu}$ are
affiliated to the centre $\mathcal Z= C_b(\Sigma)$ of the multiplier
algebra $M(\mathcal E)$ of~$\mathcal E$. 

In less technical terms this means that, given a function $f$ on $\mathbb R^4$,
a function $f(q)$ on quantum spacetime can be defined as
an element of $M(\mathcal E)$ by a generalized Weyl correspondence. 
The product of two such elements of $M(\mathcal E)$ is given by the twisted
convolution product
\[
f(q)\,g(q) 
\;=\;
(2\pi)^{-8}\int dk_1 dk_2 \,\hat f (k_1)
 \,\hat g(k_2) \,e^{-\frac{i}{2} k_1 Q k_2}
 \, e^{-i(k_1+k_2)q}\ .
\]
Here, $\hat{\phantom{t}}$ indicates the Fourier transform of a function
on $\mathbb R^4$ and $k_1, k_2$ are elements of the ordinary
Minkowski space. The exponential $\exp({-\frac{i}{2} k_1 Q k_2})$ is referred to as
the twisting. In analogy, the free field $\phi(q)$ on quantum spacetime was
formally given in~\cite{dfr} as $\qphi(q)=(2\pi)^{-4}\int dk\;\hat \cphi(k) 
\,e^{-ikq}$ where $\cphi$ is the free field on Minkowski space.

Different definitions of perturbative quantum field theory on noncommutative
spacetimes have been discussed in the literature  (cf.
e.g.~\cite{bdfp}). While these approaches are equivalent on the ordinary 
Minkowski space, they cease to be so on noncommutative spacetimes with
noncommuting time variable.

One of the possible approaches is based on what is known as the
Yang-Feldman approach  in ordinary quantum field theory. 
As early as 1952, this approach was already employed in the context of theories
with nonlocal interactions~\cite{KM}. 
%%%
Here, the field equation is the starting point, which for a self-interacting
bosonic field on the noncommutative Minkowski space may be given as follows 
\[
(\square_q + m^2)\phi(q)=-g\phi^{n-1}(q)
\]
with derivatives $\partial_{q^\mu}$ defined as the infinitesimal
generators of translations (see~\cite{dfr}). The field equation 
is then solved recursively in terms of a formal power series in the coupling
constant,
\begin{equation}\label{prodYF}
\phi=\sum_{\kappa=0}^\infty g^\kappa \,\phi_\kappa\,,
\;\;\mbox{ with }\; \phi_\kappa(q)=
\int d^4y\,G (y)\hspace{-1ex}\sum_{\sum\limits_{i=1}^{n-1}\!\kappa_i=\kappa-1}
\hspace{-2ex}\phi_{\kappa_1}(q-y)\dots\phi_{\kappa_{n-1}} (q-y)\,,
\end{equation}
$y\in \mathbb R^4$, 
where $G$ is one of the Green functions of the {\em ordinary} 
Klein-Gordon operator, chosen according to the given boundary conditions.

Unlike the modified Feynman rules~\cite{filk} which are widely used for
perturbative calculations on the noncommutative Minkowski space, neither the
Yang-Feldman approach nor the Hamiltonian approach proposed in~\cite{dfr}
entail a formal (i.e. before renormalization) violation of
unitarity even for noncommuting time variable (see~\cite{bdfp}).

Already a second order calculation performed in~\cite{bdfp} showed that the
perturbation theory in  these approaches is not free of ultraviolet
divergences. As in ordinary quantum field theory this can be traced to the fact
that products of fields  $\phi^n(q)$ are ill-defined.  Mimicking the
renormalization procedure (in position space) of ordinary quantum field theory,
the first aim thus should be to find well-defined products of fields. One of
the conceptual problems we are faced with here is to find an adequate
generalization of the locality principle on which the definition of such
products on the ordinary Minkowski space is founded. Various approaches to
address this question are possible.

In~\cite{bdfp2},  which was based on the doctoral thesis of one of the
authors~\cite{piacdiss}, we used the best localized states introduced
in~\cite{dfr} to   replace the ordinary concept of locality by a notion of
``approximate coincidence'', compatible with the uncertainty relations. The
limit of coinciding points, which usually entails the appearance of ultraviolet
divergent expressions, is replaced by the evaluation of a conditional
expectation, given by the so-called quantum diagonal map, which minimizes the
difference variables while leaving the mean coordinates invariant. Employing
this concept of approximate coincidence in the definition of the interaction
term leads to a natural regularization in quantum field theory on the
noncommutative Minkowski spacetime. No ultraviolet divergences appear. 
Unfortunately, only translation and rotation invariance are preserved in this
approach, and the free theory is treated on a different footing than the
interaction. 

In the present paper we follow a different idea.  Heuristically, a local
functional of a field is an element of the algebra  generated by the field and
its derivatives. The obstruction that, as on commutative spacetime, the  field
is too singular for admitting  pointwise products, is circumvented by smearing
the field over translations,
\[
   \qphi_{g}(q)=\int dx\, g(x)\qphi(q+x) 
\]
with a test function $g$.
The smeared fields $\qphi_g(q)$ are then well defined elements of a 
topological algebra which depend continuously on the test functions. 
We are therefore led to algebra-valued distributions\label{fieldheur}
\[
   \qphi^n_g(q)=\int dx_1 \cdots dx_n \, \qphi(q+x_1)\cdots 
               \qphi(q+x_n) \, g(x_1,\ldots,x_n) \ .
\]
Now let ${\cal O}$ be a neighbourhood of the origin of Minkowski space. 
We call $\qphi^n_g(q)$  local of order ${\cal O}$ if 
${\rm supp}g \subset {\cal O}^n$. Our aim is 
to find suitable subtractions 
\[
   \qphi_g^n-\sum_{k=1}^n\qphi^{n-k}_{\gamma_{k}^{(n)}(g)}
\]
with continuous linear maps $\gamma_k^{(n)}$ from test functions with 
$n$ variables to test functions with $n-k$ variables   such that the limit
$g\to \delta$ (limit of coinciding points) is a {\em well defined} 
quantum field on  quantum spacetime which is {\em local} of all orders.

The crucial fact now is that the usual Wick ordering is not of this type when
applied to fields on quantum spacetime, as some of the subtracted terms are not
local. We would therefore like to refrain from subtracting them and therefore 
introduce a modified Wick product, the so-called  {\em quasiplanar  Wick
product}, which is obtained by admitting only such maps $\gamma_k^{(n)}$ in 
the subtraction procedure which do not decrease the order of locality.
 Fortunately, the terms which  remain unsubtracted compared to the
ordinary Wick product turn out to be finite in the limit of coinciding points
such that our procedure yields a well-defined product in this case.

We then postulate that only quasiplanar Wick products are admissible as
counterterms in  perturbative renormalization. While this  seems to be
necessary from the point of view of locality  (and, as far as we checked up to
now, also sufficient for the absorption   of ultraviolet divergences) it
seriously modifies the asymptotic  behaviour of the theory. It turns out that
in the Yang Feldman approach  the asymptotic outgoing and  incoming free fields
are neither local nor  Lorentz-invariant, although the 
subtraction procedure itself is fully Lorentz-covariant.  We  find that the notorious
infrared-ultraviolet mixing shows up in our  framework not as an inconsistency
of the theory but in a drastic change  of the dispersion relation which we
compute  to first order in $\phi^4$-theory. This may allow new tests of the
theory.

It is noteworthy that the formalism presented here may   formally also
be applied in the Hamiltonian approach.  

It should be stressed that   
in our setting the Planck length \(\lambda_P\)
is kept fixed at its physical value.
If  one adopts the point of view that  in the limit
``$\lambda_P\rightarrow 0$'' the theory should reduce to the usual
renormalized theory on Minkowski space, one has to find additional
counterterms, which for $\lambda_P\neq 0$   correspond to finite
renormalizations and in the limit ``\(\lambda_P\rightarrow 0\)'' produce the
missing ordinary counterterms needed on Minkowski space. So far, we have not
been able to find a  local and  Lorentz invariant definition of such
counterterms.

Also in view of the modified dispersion relation, it seems that in all our
attempts to introduce interactions of fields on quantum spacetime, Lorentz
invariance is sooner or later lost -- although the underlying geometry of our
model of quantum spacetime as well as the  theory of free fields on quantum
spacetime are fully Lorentz (and Poincar\'e) invariant. This point calls for a
deeper understanding we still lack at the moment.

We would like to emphasize that results
regarding the renormalization of field theories on a
noncommutative Euclidean spacetime~\cite{roiban} cannot be directly 
applied to field theories on the noncommutative  Minkowski space. We will see
explicitly in an example that a tadpole which is finite in the Euclidean
setting fails to be so on the noncommutative Minkowski spacetime. This is not
very surprising as no generalization of Osterwalder-Schrader positivity seems
to be available and not even the Wick rotation  itself  has been given
proper meaning in a space/time noncommutative setting. 

We will furthermore see that a theory of self-interacting scalar fields with
commuting time variable cannot be renormalized by local counterterms.

This paper  focuses on the combinatorial aspects and the physical
consequences of the idea to admit only local counterterms.  The  full proof 
 that quasiplanar Wick products are well defined at coinciding points
($g\rightarrow \delta$) turned
out to be rather technical and is merely sketched in this paper. 
Details regarding domains of definition and appropriate test function
spaces will be subject of a forthcoming publication.

The results presented here are based to a large extent on the doctoral 
thesis of one of the authors \cite{bahnsdiss} where further details 
may be found.

%%% 

%%%%%%%%%%%%%%%%%%%%%%%%%%%%%%%%%%%%%%%%%%%%%%%%
%%%%%%%%%%%%%%%%%%%%%%%%%%%%%%%%%%%%%%%%%%%%%%%%

\section{Fields on the noncommutative Minkowski space}

In \cite{dfr}, the quantization of a function $f(x)$ on ordinary
spacetime was defined in terms of the Weyl correspondence 
\[
W(f)\equiv f(q):= \int dk\; e^{ikq}\check f(k) = (2\pi)^{-4} \int dk \; 
e^{-ikq}\hat f(k) \ ,
\]
where $\check f(k)=(2\pi)^{-4}\int dx\; f(x)e^{-ikx}$, 
$\hat f(k)=(2\pi)^{4}\,\check f(-k)$. By analogy, the free 
field $\qphi$ on the quantum spacetime was defined by the heuristic 
formula 
\[
\qphi (q)
=(2\pi)^{-4}\int dk\;\hat \cphi(k) \otimes e^{-ikq}\ ,
\]
where $\cphi$ is the free field on Minkowski space and $\hat \cphi$ is its
Fourier transform. $\qphi(q)$  is  to be thought of as a (formal)  element of
the tensor product $\mathfrak F\otimes\mathscr E$, where  $\mathfrak F$ is the
algebra of polynomials of the free field. 
Roughly speaking, this means that after evaluation in a suitable state $\omega$
on $\mathcal E$, we obtain an element of $\mathfrak F$. A precise definition
can be given in terms  of the dual $W^*$ of the Weyl quantization (known as the
Wigner transform), which is defined by
\[
%\hat\psi_\omega(k)\equiv \widehat{W^*\omega}(k)=\omega(e^{ikq}) \ ,
\check\psi_\omega(k)\equiv 
%(2\pi)^{-4}\,\widehat{W^*\omega}(-k)
\invF{W^*\omega}(k)=(2\pi)^{-4}\,\omega(e^{-ikq}) \ ,
\]
where $\omega$ is a state on $\mathscr E$.  Note  that $k\mapsto
\omega(e^{ikq})$ defines a function in the Schwartz space 
$\mathcal S(\mathbb R^4)$, provided that \(\omega\) is in the domain
of all monomials in the \(q^\mu\)'s (since 
\(\tfrac{\partial}{\partial\alpha_\mu}e^{i\alpha q}=ie^{i\alpha q}
(q^\mu+\tfrac{1}{2}(Q\alpha)^\mu)\), and  \(i\alpha_\mu e^{i\alpha q}=
[(Q^{-1}q)_\mu,e^{i\alpha q}]\)). For such \(\omega\) we may set
\[
\qphi(\omega)\equiv(W\cphi)(\omega):= \cphi(W^*\omega)\ , 
\]
and with this definition,  
a quantum field on quantum spacetime is an affine functional
on a suitable $*$-weakly dense subset of the state space 
$S(\mathscr E)$ of $\mathscr E$,
taking values in $\mathfrak F$.
In this sense, we may now write
\[
\qphi(\omega) =
\int dx\,\cphi(x)\,\psi_\omega(x)=\int dk\,\hat \cphi(k) \check \psi_\omega(k)\
,
\]
and thus recover an expression which is well known from field theory
on Minkowski space.

The positivity property of the state $\omega$ implies that the field 
$\qphi$ respects the Heisenberg uncertainty relations for the simultaneous 
determination of the coordinates. Nevertheless, the field is still too singular
to admit (pointwise) products: indeed,
\[
(k_1,k_2)\mapsto \check \psi_{\omega}^{(2)}(k_1,k_2)\equiv
(2\pi)^{-8}\,\omega(e^{-ik_1q}e^{-ik_2q})
\]
fails to be strongly decreasing.

Therefore, as mentioned in the introduction, we smear the quantum field 
over translations. Let $f\in\mathcal S(\mathbb R^4)$. Then we set
\[
\qphi_f(\omega)\equiv\cphi(\psi_\omega\times f) \ ,
\]
where $\times$ denotes the ordinary convolution product; eventually, we will be
interested in the limit $f \rightarrow \delta$. According to the above
discussion, $\qphi_f(\omega)$ can be written as 
\[
\qphi_f(\omega)=\int dx\,\cphi(x)(\psi_\omega\times f)(x)
= \int dk\, \hat \cphi(k)\,\check f(k)\,(2\pi)^{4} \,\check
\psi_\omega(k)\ ,
\]
and in order to establish the connection with the heuristic formula on
page~\pageref{fieldheur}, we note that formally, this can be understood as the
evaluation of 
\[
\int dx \,\qphi(q+x)\,f(x) = 
\int dk \,\hat \cphi(k)\,\check f(k)\otimes 
e^{-ikq}
\]
in a state $\omega$, since $\omega(e^{-ikq})=(2\pi)^{4} \,\check
\psi_\omega(k)$ by definition. 

Now, the $n^{\text{th}}$ power of $\qphi_f$ exists and is given by
\[
(\qphi_{f})^n(\omega)=\cphi^{\otimes n}(\psi^{(n)}_\omega\times 
f^{\otimes n}) \ ,
\]
with $f\in\mathcal S(\mathbb R^{4})$, 
\begin{equation}\label{psiom}
\check\psi_\omega^{(n)}(k_1,\dots,k_n)=(2\pi)^{-4n}\,
\omega(e^{-ik_1q}\dots e^{-ik_nq})\ ,
\end{equation} 
and where $\cphi^{\otimes n}$ is the operator
valued distribution
\[
\cphi^{\otimes 
n}(x_{1},\ldots,x_{n})=\cphi(x_{1})\cdots\cphi(x_{n}) \ .
\]
More generally, for
$f\in\mathcal S(\mathbb R^{4n})$, we may define regularized products of fields
by 
\[
\qphi^{n}_f(\omega)=\cphi^{\otimes n}(\psi^{(n)}_\omega\times f) \ ,
\]
so that
\[
(\qphi_{f})^n=\qphi^{n}_{f^{\otimes n}} \ .
\]
Products of regularized fields are defined by 
\[
\qphi^{n}_f\;\qphi^{m}_g=\qphi^{n+m}_{f\otimes g},\quad
f\in\mathcal S(\mathbb R^{4n})\ , \ g\in\mathcal S(\mathbb R^{4m}) \ ,
\]
and the adjoint is given by
\[
{\qphi_f^n}^*={\qphi_{f^*}^n} \ ,
\]
where $f^*(x_1,\dots,x_n)=\overline{f(x_n,\dots,x_1)}$.

Given a regular representation of $\mathscr E$ on some Hilbert space  $\mathscr
H$, the (formal) elements $\qphi_f^{n}$ of $\mathfrak F\otimes\mathscr E$ can
be represented by operators on a dense domain in  $\mathscr
H_\cphi\otimes\mathscr H$, where $\mathscr H_\cphi$ is the Fock space of the
free field. 

%%%%%%%%%%%%%%%%%%%%%%%%%%

%We now want to define products of fields which remain
%welldefined in the limit where the regularization is removed, i.e. for
%$f\rightarrow \delta$. We are thence looking for suitably subtracted products

We now look for suitably subtracted products of fields
\label{suitdef}
\[
\wickq{\qphi_f^n}\;=\;\sum_{k=0}^n\qphi_{\gamma_k^{(n)}(f)}^{n-k}\;=\;
\qphi_f^{n}+\sum_{k=1}^n\qphi_{\gamma_k^{(n)}(f)}^{n-k}
\]
where $\gamma_k^{(n)}:\mathcal S(\mathbb R^{4n})\rightarrow
\mathcal S(\mathbb R^{4(n-k)})$, $k=0,\ldots,n$, 
are continuous linear maps, such that
\begin{enumerate}
\item  \label{coincpts} when $f\rightarrow\delta$, 
the limit of $\wickq{\qphi_f^n}$ exists as an affine
$\mathfrak F$-valued functional on some dense subset of 
$S(\mathscr E)$;
\item \label{loc} the maps $\gamma_k^{(n)}$ can be chosen to be local
in the sense that 
\[
\text{supp}\gamma_k^{(n)}(f)\subset
\overline{\bigcup_{\substack{U\subset\{1,\dots,n\}\\|U|=n-k}}
P_U\text{supp}f},
\]
where $P_U$ is the projection $\mathbb R^{4n}\mapsto\mathbb R^{4|U|}$
given by
\[
P_U(x_1,\dots,x_n)=(x_u)_{u\in U}.
\]
\end{enumerate}
Note that condition 2. ensures that in the limit where $f\rightarrow\delta$, 
the product of fields (if it exists) is local of all orders.

%%%%%%%%%%%%%%%%%%%%%%%%%%%%

In order to clarify the above idea, let us first  discuss the ordinary Wick
product $\wickn{\cphi^{\otimes n}}$ on Minkowski space. It is obtained from the
product $\cphi^{\otimes n}$ by ``putting all annihilation operators to the
right'', or equivalently, given by an alternating sum over all possible
contractions of $n$ fields. To put this latter definition into a compact form,
we now introduce the following notation.

%%%%%%%%%%%%%%%%%%%%%%%%%%

Let $N$ be a finite ordered set. A contraction in $N$ is a  pair consisting of
a subset $A\subset N$ and an injective map $\alpha:A\to  N\setminus A$ such
that $\alpha(a)> a$ for all $a\in A$ (with respect to the order of $N$).  The
set of  all contractions in $N$, including the empty contraction with
$A=\emptyset$, is denoted by $\mathscr C(N)$. $A$ is  considered as an ordered
subset of $N$ (with its natural order) and  $\alpha(A)$ is an ordered set which
inherits its order  $\stackrel{ \alpha}{<}$ from $A$ via the map $\alpha$
(i.e.  $\alpha(a)\stackrel{ \alpha}{<}\alpha(a')$ if $a<a'$). In what follows,
the letter $U$ will denote the set of  uncontracted indices,
$U=N\setminus(A\cup\alpha(A))$. If different  contractions $C$ are involved, we
label $A,\alpha,U$ by a lower index $C$.

To every contraction $C\in \mathscr C(N)$ we associate a linear 
continuous map,
the so-called contraction map 
\[
\gamma_{0}^{C}:\mathcal S(\mathbb R^{4|N|}) \to \mathcal S(\mathbb 
R^{4|U|})\ ,
\]
by 
\[
\gamma_{0}^{C}(f)(x_{U}) = \int dx_{A}dx_{\alpha(A)} 
\prod_{a\in A}\Delta_{+}(x_{a}-x_{\alpha(a)}) f(x_{N}) \ .
\]
Here, $\Delta_+$ denotes  the ordinary $2$-point function of the free
field and we have used the convention that, for a finite ordered set $B$, 
$x_{B}$ denotes the tuple $x_{B}=(x_{b_{1}},\ldots,x_{b_{|B|}})$ with 
$b_{1}<b_{2}<\cdots < b_{|B|}$.

In Fourier space, the contraction map assumes the form 
\[
\invF{\gamma_{0}^{C}(f)}(k_{U}) = (2\pi)^{8|A|}
\int d\mu_{A}(k_{A}) 
\check{f}(k_{N})\big|_{k_{\alpha(A)}=-k_{A}}\ ,
\]
where $d\mu_{A}(k_{A}) = \prod_{a\in A}d\mu(k_{a})$ with $d\mu(k)$ denoting 
the Lorentz-invariant measure on the mass shell
\[
d\mu(k)= (2\pi)^{-3}\frac{d{\bf k}}{2\omega_{\bf 
k}}\Big|_{k_{0}=\omega_{\bf k}}\ , \quad
\omega_{\bf k}=\sqrt{m^2+{\bf k}^2}\ .
\]
Making use of the contraction maps $\gamma_{0}^{C}$, we can now write the
Wick products  on commutative spacetime as the alternating sum
\[%begin{equation}\label{ordWick}
\wickn{\cphi^{\otimes |N|}} = \sum_{C\in\mathscr C(N)}(-1)^{|A|} 
\cphi^{\otimes |U|}\circ \gamma_{0}^{C} \ ,
\]%end{equation}
where for $U=\emptyset$, $\cphi^{\otimes |U|}=1$.
Roughly speaking, for coinciding arguments, the right hand side of the above
consists of a vertex with $n$ legs plus (or minus) all possible tadpoles.

%%%%%%%%%%%%%%%%%%%%%%%%%%

A first attempt to define suitably subtracted products of fields on the
noncommutative Minkowski space was to generalize the ordinary Wick products to
the noncommutative spacetime~\cite{dfr}. However, while this prescription
fulfills condition 1, it violates condition 2, as we shall see below.

Before proceeding, we observe that any state $\tilde \omega \in \mathcal
S(\mathcal E)$ can be decomposed as $\tilde \omega = \mu\circ \omega$, where
$\mu$ is a probability measure on $\Sigma$ and
$\omega$ is a positive, unital, $\mathcal Z$-linear map taking values in
$\overline{\mathcal Z}=L^\infty(\Sigma,\mu)$ (a ``$\overline{\mathcal
Z}$-valued state'') with   
\label{statedec}
\[
\omega(\prod_{j\in N} e^{ik_jq})=  e^{-\frac i 2 \sum_{j<l} k_j Q
k_l}\,\omega(e^{i\sum_{j\in N} k_jq}) \quad \in \;
\overline{\mathcal Z}
\ .
\]
Unfortunately, no
Lorentz-invariant choice of $\mu$ exists. Particular choices of
$\mu$ are the measure which is supported on the rotation and translation
invariant subset $\Sigma_1\subset \Sigma$ (see~\cite{dfr}) and the point
measure. The latter choice can equivalently be understood as the case where a
fixed noncommutativity matrix $[q^\mu,q^\nu]=i \theta^{\mu\nu}\; \in\Sigma$ is
used and $\mathcal Z$ is trivial. This special case is therefore included in our
more general setting. In the considerations which follow, the
integration over $\Sigma$ will for the most part be irrelevant, and we
therefore refrain from performing it until the very last. Note that the
formalism is fully covariant, but that we will  frequently replace the
operators $Q^{\mu\nu}$ by generic spectral values $\sigma^{\mu\nu}$,
$\sigma\in \Sigma$, in the sense of the joint functional calculus of the
$Q^{\mu\nu}$.  
If necessary, we will furthermore consider $\overline{\mathcal Z}$-valued test
functions, distributions, Hilbert space vectors and operators.

%%%%%%%%%%%%%%%%%%%%%%%%%

We now set 
\[
\wickn{\qphi_{f}^n}(\omega) = \wickn{\cphi^{\otimes 
n}}(\psi_{\omega}^{(n)}\times f)\ ,
\]
with $f \in \mathcal S(\mathbb R^{4n})$ and with $\invF{{\psi_{\omega}^{(n)}}}$
given by (\ref{psiom}).
%\begin{equation}\label{psiom}
%\invF{{\psi_{\omega}^{(n)}}} (k_{1},\dots, k_n)=
%(2\pi)^{-4n}e^{-\frac i 2\sum
%k_j Q k_l} \,\omega(e^{-i(k_1+\cdots +k_{n})q})\ .
%\end{equation}
From the above it then follows that 
\[
\wickn{\qphi_{f}^{|N|}}(\omega)=\wickn{\cphi^{
{\otimes |N|}}}(\psi_{\omega}^{(|N|)}\times f) 
=\sum_{C\in\mathscr C(N)}(-1)^{|A|} 
\cphi^{\otimes |U|}\big(\,\gamma_{0}^{C}(\psi_{\omega}^{(|N|)}\times f)
\,\big) \ .
\]
We now define the {\em quantum contraction} $\gamma^{C}$ by
requiring (for $\overline{\mathcal Z}$-valued states $\omega$)
\[
\gamma_{0}^{C}(\psi_{\omega}^{(|N|)}\times f)= 
\psi_{\omega}^{(|U|)}\times \gamma^{C}(f) \ ,
\]
such that
\[%begin{equation}\label{Winc}
\wickn{\qphi_{f}^{|N|}}(\omega) = \sum_{C\in\mathscr
C(N)}(-1)^{|A|} \; \qphi^{ |U|}_{\gamma^{C}(f)}(\omega)\ .
\]

To compute $\gamma^{C}$ we use the fact 
that due to the commutation relations of coordinates on quantum 
spacetime we have  
\[
\ts{\prod\limits_{j\in N}} e^{-ik_jq}
\big|_{k_{\alpha(A)}=-k_{A}} = 
\ts{\prod\limits_{j\in U}} e^{-ik_jq}\,
e^{-i\langle k_{A},I k_{A}\rangle -i \langle k_{A},E k_{U}\rangle}
\]
where $I$ is a $4|A|\times 4|A|$ matrix (called the {\em intersection  matrix})
and $E$ a $4|A|\times 4|U|$ matrix (called the {\em enclosure  matrix}) with
$4\times 4$ blocks, where  (with respect to the natural order of both $A$
and $\alpha(A)$ as subsets of $N$),
\beqa
I_{aa'}&=&\left\{ 
\begin{array}{ccc}
    Q & , & \text{if } a<a'<\alpha(a)<\alpha(a')  \\
    0 & , & \text{otherwise}
\end{array}
\right.
\\
E_{au}&=&\left\{ 
\begin{array}{ccc}
    Q & , & \text{if } a<u<\alpha(a) \\
    0 & , & \text{otherwise}
\end{array}
\right.
\eeqa
and where for two momenta $k,k'$ the contraction with $Q$ is defined 
by 
$kQk'=k_{\mu}Q^{\mu\nu}k'_{\nu}$.

We thus obtain 
\begin{equation}\label{gammaC}
\invF{\gamma^{C}(f)}(k_{U})= (2\pi)^{8|A|}  \int d\mu_{A}(k_{A})
e^{-i\langle k_{A},I k_{A}\rangle -i\langle k_{A},E 
k_{U}\rangle}\, \check f(k_N)|_{k_{\alpha(A)}=-k_{A}} \ .
\end{equation}

In terms of graphs, these definitions can be visualized as follows:  For the
ordered set $N=(1,\dots,n)$ draw a number of $n$ points in a horizontal line.
For a contraction $C$, connect each point $a\in A$ with its respective  partner
$\alpha(a)$  by a curve in the upper half  plane (called an {\em internal
line}). 
\label{graphs}
Then the entry  $I_{aa'}$ of the intersection matrix is nonzero if and
only if their connecting curves intersect and $a<a'$, and the  entry $E_{au}$
of the enclosure matrix vanishes if and only if the vertical line from $u$ to
$+\infty$  (called an {\em external line}) crosses the internal line connecting
$a$  and $\alpha(a)$.

\example{Consider the contraction $C$ in $N=(1,\dots,8)$ where 
$A=(2,4,6)$ and $\alpha(A)=(3,7,8)$. The corresponding graph then is
\begin{picture}(80,10)(-5,0)
\put(0,2){\circle{2}}
\put(10,2){\circle{2}}
\put(20,2){\circle{2}}
\put(30,2){\circle{2}}
\put(40,2){\circle{2}}
\put(50,2){\circle{2}}
\put(60,2){\circle{2}}
\put(70,2){\circle{2}}
\put(0,-3){\tiny{1}}
\put(10,-3){\tiny{2}}
\put(20,-3){\tiny{3}}
\put(30,-3){\tiny{4}}
\put(40,-3){\tiny{5}}
\put(50,-3){\tiny{6}}
\put(60,-3){\tiny{7}}
\put(70,-3){\tiny{8}}
\qbezier(10,2)(15,13)(20,2)
\qbezier(30,2)(45,13)(60,2)
\qbezier(50,2)(60,13)(70,2)
\end{picture},
and it allows  to directly read off the intersection and the enclosure matrix: 
$I_{46}=Q$,
$E_{45}=Q$, all others 0.}

Note moreover that every contraction may be naturally decomposed into connected
  com\-po\-nents as is illustrated by the following example.

\example{The contraction $C\in\mathscr C(\{1,\ldots,9\})$ where 
$A=(1,2,4,5)$, and
$\alpha(A)=(9,7,6,8)$ has two connected components $C_{1}$ and $C_{2}$
with $A_{C_{1}}=\{1\}$ and $A_{C_{2}}=\{2,4,5\}$. In terms of graphs, 
the connected components of 
\[
\begin{picture}(110,20)(-5,-5)
\put(0,2){\circle{2}}
\put(10,2){\circle{2}}
\put(20,2){\circle{2}}
\put(30,2){\circle{2}}
\put(40,2){\circle{2}}
\put(50,2){\circle{2}}
\put(60,2){\circle{2}}
\put(70,2){\circle{2}}
\put(80,2){\circle{2}}
\put(0,-3){\tiny{1}}
\put(10,-3){\tiny{2}}
\put(20,-3){\tiny{3}}
\put(30,-3){\tiny{4}}
\put(40,-3){\tiny{5}}
\put(50,-3){\tiny{6}}
\put(60,-3){\tiny{7}}
\put(70,-3){\tiny{8}}
\put(80,-3){\tiny{9}}
\qbezier(0,2)(40,30)(80,2)
\qbezier(10,2)(35,20)(60,2)
\qbezier(30,2)(40,13)(50,2)
\qbezier(40,2)(55,13)(70,2)
\end{picture}
\]
are given by
\begin{picture}(95,10)(-5,-2.5)
\put(0,2){\circle{2}}
\put(10,2){\circle{2}}
\put(10,2){\circle{2}}
\put(20,2){\circle{2}}
\put(30,2){\circle{2}}
\put(40,2){\circle{2}}
\put(50,2){\circle{2}}
\put(60,2){\circle{2}}
\put(70,2){\circle{2}}
\put(80,2){\circle{2}}
\put(0,-3){\tiny{1}}
\put(10,-3){\tiny{2}}
\put(20,-3){\tiny{3}}
\put(30,-3){\tiny{4}}
\put(40,-3){\tiny{5}}
\put(50,-3){\tiny{6}}
\put(60,-3){\tiny{7}}
\put(70,-3){\tiny{8}}
\put(80,-3){\tiny{9}}
\qbezier(0,2)(40,20)(80,2)
\end{picture}
($C_{1}$)~~and
\begin{picture}(95,10)(-5,-2.5)
\put(0,2){\circle{2}}
\put(10,2){\circle{2}}
\put(20,2){\circle{2}}
\put(30,2){\circle{2}}
\put(40,2){\circle{2}}
\put(50,2){\circle{2}}
\put(60,2){\circle{2}}
\put(70,2){\circle{2}}
\put(80,2){\circle{2}}
\put(0,-3){\tiny{1}}
\put(10,-3){\tiny{2}}
\put(20,-3){\tiny{3}}
\put(30,-3){\tiny{4}}
\put(40,-3){\tiny{5}}
\put(50,-3){\tiny{6}}
\put(60,-3){\tiny{7}}
\put(70,-3){\tiny{8}}
\put(80,-3){\tiny{9}}
\qbezier(10,2)(30,20)(60,2)
\qbezier(30,2)(40,13)(50,2)
\qbezier(40,2)(50,13)(70,2)
\end{picture} ($C_2$).
}

%%%%%%%%%%%%%%%%%%%%%%%%%%%%%%
\section{Quasiplanar Wick products}
%%%%%%%%%%%%%%%%%%%%%%%%%%%%%%

According to the programme outlined on page~\pageref{suitdef}  we now want to
introduce subtracted products of fields on the noncommutative Minkowski space
which are defined in  terms of local contractions only. This condition is not
satisfied by ordinary Wick products. To see this, consider the third Wick
power $\wickn{\qphi_{f}^{ 3}} $ which in terms of graphs is given by the
following sum of contractions 
$\;\begin{picture}(25,10)
\put(0,2){\circle{2}}
\put(10,2){\circle{2}}
\put(20,2){\circle{2}}
\end{picture}
-\;\;\begin{picture}(25,10)
\put(0,2){\circle{2}}
\put(10,2){\circle{2}}
\put(20,2){\circle{2}}
\qbezier(0,2)(5,12)(10,2)
\end{picture}
-\;\begin{picture}(25,10)
\put(0,2){\circle{2}}
\put(10,2){\circle{2}}
\put(20,2){\circle{2}}
\qbezier(10,2)(15,12)(20,2)
\end{picture}
-\;\begin{picture}(25,10)
\put(0,2){\circle{2}}
\put(10,2){\circle{2}}
\put(20,2){\circle{2}}
\qbezier(0,2)(10,15)(20,2)
\end{picture}
$. 
The last contraction yields
\[
\gamma^{C}(f)(x_2)=
\int dx_1 dx_3 \int \! d\mu(  k) \,e^{-ik(x_1-x_3)}
\,f (x_1, x_2+Qk,x_3)
\]
where we have performed the fibrewise-defined coordinate transformation $x_2
\rightarrow x_2 + \sigma k$. This expression clearly violates the locality
condition (condition~\ref{loc} on page~\pageref{suitdef}). For
$f(x_1,x_2,x_3)=\delta(x_1-x_2)\delta(x_1-x_3)g(x_1)$ (which renders a
well-defined expression, as we shall see below) it was shown in~\cite{bahnsdiss}
that this nonlocality cannot be cured by adding a correction term from the
range of the Klein-Gordon operator.

It is easy to see that a contraction is local if 
its enclosure matrix vanishes, since in 
this case the uncontracted variables decouple from the contracted 
variables and we find 
\[
\text{supp }\gamma^{C}(f)\subset \overline{P_{U}\text{supp } f}\ .
\]
The contractions with vanishing  enclosure matrix may be represented by graphs
whose external lines are not crossed by internal lines. We call
these graphs (and  the corresponding contractions) quasiplanar. The
set of  contractions for which {\em all} connected components are quasiplanar 
will be denoted by $\mathscr C_{qp}(N)$. 
Note that due to the definition of connected components used here (which
differs from the one in~\cite{bahnsdiss} and simplifies the combinatorics
below), 
the contraction 
\begin{picture}(60,10)(-5,0)
\put(0,2){\circle{2}}
\put(10,2){\circle{2}}
\put(20,2){\circle{2}}
\put(30,2){\circle{2}}
\put(40,2){\circle{2}}
\put(50,2){\circle{2}}
\qbezier(0,2)(25,20)(50,2)
\qbezier(10,2)(20,13)(30,2)
\qbezier(20,2)(30,13)(40,2)
\end{picture}
is quasiplanar but not in $\mathscr C_{qp}(N)$.
%%%
 
We now define  the quasiplanar Wick products by the following formula
($f\in\mathcal S(\mathbb R^{4|N|})$)
\begin{equation}\label{qpWick}
\wickq{\qphi^{|N|}_{f}} = \sum_{C\in\mathscr 
C_{qp}(N)}(-1)^{\kappa}\qphi^{|U|}_{\gamma^{C}(f)} \ ,
\end{equation}
where $\kappa$ is the number of connected components of $C$. For an example 
see appendix~\ref{exqpWick}. It is clear that by definition quasiplanar Wick
products fulfill the locality condition.

With the initial conditions $\wickq{1}=1$ and 
$\wickq{\qphi}=\qphi$, the quasiplanar Wick products can be uniquely 
characterized by the recursion relation ($f\in\mathcal S(\mathbb  R^4),
g\in\mathcal S(\mathbb R^{4|N|})$)
\begin{equation}\label{qprecursion}
\wickq{\qphi^{|\{1\}\sqcup N|}_{f\otimes 
g}}=\qphi_{f}\; \wickq{\qphi^{|N|}_{g}} - \sum_{\substack{C\in\mathscr 
C_{qp}(\{1\}\sqcup N)\\ C\text{ connected} \\ 1\in A}} 
\wickq{\qphi^{|U|}_{\gamma^{C}(f\otimes g)}} \ .
\end{equation}
Here, the symbol $\sqcup$ denotes the disjoint union of two ordered 
sets, where the second set is appended to the first set, such that for all
$n\in N, m\in M$, $n<m$ in $N\sqcup M$. For an example of (\ref{qprecursion})
see appendix~\ref{exqprecursion}.

Instead of directly proving the recursion relation~(\ref{qprecursion}),
we prove the
analogue of Wick's theorem of which (\ref{qprecursion}) is a corollary. Let $N$
and $M$ be ordered finite sets. We let $\mathscr  C(N,M)$ denote the set of all
quasiplanar contractions $C\in \mathscr  C(N\sqcup M)$ which have the  property
that every connected component $C'$ of $C$ connects $N$ and $M$, in  the sense
that $\alpha( A_{C'}\cap N)\cap M\neq\emptyset$.  
Note that $C\in \mathscr  C(N, M)$ is in general not in $\mathscr 
C_{qp}(N\sqcup M)$.

\begin{thm} {\bf ``Wick's theorem for Quasiplanar Wick products'':}
Let $f\in\mathcal S(\mathbb R^{4n})$ and $g\in\mathcal S(\mathbb 
R^{4m})$. Then
\begin{equation}\label{qpWickthm}
\wickq{\qphi^n_{f}}\; \wickq{\qphi^m_{g}} = \sum_{C\in\mathscr 
C(N,M)} \wickq{\qphi^{|U|}_{\gamma^{C}(f\otimes g)}}
\end{equation}
where $N=\{1,\ldots,n\}$ and $M=\{n+1,\ldots,n+m\}$
\end{thm} 
\proof{After inserting the definition of quasiplanar Wick products
(\ref{qpWick}), the left hand side is
\[
\sum_{\substack{C\in\mathscr C_{qp}(N)\\C'\in\mathscr C_{qp}(M)}} 
(-1)^{\kappa_{C}+\kappa_{C'}} 
\qphi^{|U_{C}|+|U_{C'}|}_{\gamma^{C}(f)\otimes\gamma^{C'}(g)} \ .
\]
For the right hand side we find
\[
\sum_{C\in\mathscr C(N,M)}\sum_{C'\in\mathscr 
C_{qp}(U_{C})}(-1)^{\kappa_{C'}}
\qphi^{|U_{C'}|}_{\gamma^{C'}\circ\gamma^{C}(f\otimes g)} \ .
\]
In the latter expression, $C'$
may be decomposed into 3 mutually 
disconnected contractions 
$C_{1}$, $C_{2}$ and $C_{3}$ where $C_{1}\in\mathscr C_{qp}(N\cap 
U_{C})$, 
$C_{3}\in\mathscr C_{qp}(M\cap U_{C})$ and $C_{2}\in\mathscr 
C(N\cap U_{C},M\cap U_{C})$.
Note that $C_{2}$ is connected since $C^\prime \in \mathcal C_{qp}(U_C)$. 
We may now combine $C$ and $C_{2}$ to a single contraction 
$C_{4}\in\mathscr C(N,M)$. We observe that every nonempty contraction 
$C_{4}\in\mathscr C(N,M)$ appears twice in the sum, but with opposite 
signs. Hence all these contributions cancel, and only the empty 
contraction remains which yields the theorem. 
}

Two concrete applications of Wick's theorem for quasiplanar Wick products
(formula (\ref{qpWickthm})) may be found in
appendix~\ref{wickex}. 

%%%%%%%%%%%%%%%%%%

We will now give a closed formula specifying the relation  between quasiplanar
Wick products and ordinary Wick products. In fact, we show that  
quasiplanar Wick polynomials can be expressed in terms of Wick polynomials via 
the formula
\begin{equation}\label{qpWinW}
\wickq{\qphi^{|N|}_f} \;= \sum_{C\in\mathscr{C}_{ap}(N)}
                        \wickn{\qphi^{|U|}_{\gamma^C(f)}} 
		 \ .
\end{equation}
Here, $\mathscr{C}_{ap}(N)$ is the set of all {\em aplanar} contractions of $N$. 
A contraction is called aplanar if for every connected component the 
corresponding part of the enclosure matrix is nontrivial. 
Note that the empty contraction  is quasiplanar 
and aplanar, and that contractions may be neither 
in $C_{qp}$ nor in $C_{ap}$. For an example of (\ref{qpWinW}) see 
appendix~\ref{exqpWinW}.

We prove formula (\ref{qpWinW}) by showing that it satisfies the recursion
relation (\ref{qprecursion}).  The initial conditions are
obviously fulfilled. Now for the first term on the  right hand side of the
recursion relation we find, using~(\ref{qpWinW})  and Wick's theorem (for
ordinary Wick products),
\begin{equation}\label{term1}
\qphi_f \wickq{\qphi^{|N|}_g} = 
\sum_{\substack{C\in\mathscr{C}_{ap}(\{1\}\sqcup N)\\
                1\not \in A}}
                \left(\wickn{\qphi^{|U|}_{\gamma^C(f\otimes g)}} 
                      +\sum_{u\in U\setminus\{1\}}
                       \wickn{\qphi^{|U|-2}_{\gamma^{(1,u)}
                       \circ \gamma^C(f\otimes g)}} \right)
\end{equation}
where $(1,u)$ is the contraction with $A=\{1\}$ and $\alpha(1)=u$.
Applying~(\ref{qpWinW}) also to the second term in the recursion relation
yields 
\[
-\sum_{C\in\mathscr{C}(\{1\},N)}
 \sum_{C'\in\mathscr C_{ap}(U_C)}
  \wickn{\qphi^{|U_{C'}|}_{\gamma^{C'}\circ\gamma^C(f\otimes g)}}  \ .
\]
The combined contractions from~(\ref{term1}) may be decomposed
into  connected components. Now, those contractions for which the component 
containing 1 has a vanishing enclosure matrix cancel  with the second term
in the recursion relation. Hence, only the sum over all aplanar contractions  of
$\{1\}\sqcup N$ remains, which proves the claim.   

 Formula (\ref{qpWinW}) shows explicitly that the limit ``$\lambda_P
\rightarrow 0$'' does not yield the ordinary Wick products, since 
\[
\sum_{C\in\mathscr{C}_{ap}(N)}
                        \wickn{\qphi^{|U|}_{\gamma^C(f)}}
      \;=\;	\wickn{\qphi^{|N|}_{f}} \;
		        + \sum_{\substack{C\in\mathscr{C}_{ap}(N)\\U\neq N}}
                        \wickn{\qphi^{|U|}_{\gamma^C(f)}}
\]
and the terms which compared to the ordinary Wick product remain
unsubtracted do not vanish in this limit.

%%%%%%%%%%%%%%%%%%%%%%%%%%%%%%%%%%%%%%%%%%

\section{Quasiplanar Wick products at coinciding points
   (sketch)}

Let us now consider a quasiplanar Wick product  at coinciding points,
i.e. an expression of the form $\wickq{\phi^{n}_g}(q)$ where
$g(x_N)=\prod_{j=1}^n \delta (x_j)$, $N=\{1,\dots,n\}$. We will sketch
 an argument showing that such a product is well-defined.  The
mathematical details will be treated in a forthcoming publication. The proof
is based on the idea that using (\ref{qpWinW}) we may rewrite the quasiplanar
Wick product in terms of ordinary Wick products and that for a  suitable test
function $h$,  the normal ordered product of fields at coinciding points,
\[
\int dk_U\,\wickn{\prod_{i\in U} \hat \varphi (k_i)}\;
\check h(\sum_{j\in U} k_j)=\int dx \;\wickn{\cphi (x)^{|U|}} \;h(x)
\]
is a welldefined element of $\mathfrak F$.
 
We therefore apply   (\ref{qpWinW}) to $\wickq{\phi^{n}_g}$ and evaluate the
resulting expression in a suitable state $\tilde \omega=\mu\circ \omega$ to
obtain  
\begin{equation}\label{limc}
\wickq{\qphi_g^{n}}(\tilde \omega)= \sum_{C\in \mathcal C_{ap}(N)}
\mu\Big(\,\int dk_U\,\wickn{\prod_{i\in U} \,(2\pi)^{4|U|}\,\check \varphi (k_i)}\,
\invF{\gamma^{C}(g)}(k_{U})\,
\hat \psi^{(|U|)}_{\omega}(-k_U)
\,\Big) \ ,
\end{equation}
where $\invF{\gamma^{C}(g)}(k_U)$ is a bounded (not rapidly decreasing)
function of $k_U$ given by~(\ref{gammaC}) with $\check g (k_N)\equiv
(2\pi)^{-4n}$, and where 
$\check \psi^{(|U|)}_{\omega}$ which is given by (\ref{psiom})
is quickly
decreasing only in the sum of the momenta. 
Let us now pick an arbitrary contribution to the right hand side of
(\ref{limc}). Using $\cphi(x)=(2\pi)^{3/2} \int d\mu(k)\,(a(k) \,e^{-ikx} +
a^*(k) \,e^{+ikx})$, we then decompose  the Wick polynomial  into a sum of 
normal ordered products of creation and annihilation operators   $\prod_{u\in
U\setminus U'}a^{*}(k_{u})  \prod_{u'\in U'}a(k_{u'})$ with  $U'\subset U$. 

We now consider the pure creation part ($U^\prime =\emptyset$), since we know
from ordinary field theory that it is the term requiring most care in a
product of fields at coinciding points. From (\ref{gammaC}) we conclude that in
this case, $\invF{\gamma^{C}(g)}(k_U)$ is of the form
\[
(2\pi)^{-4 |U|}\int d\mu_{A}(k_{A})
e^{-i\langle k_{A},Ik_{A}\rangle -i\langle k_{A},E  (-k_{U})\rangle} 
\ ,
\]
where all momenta are on the {\em positive} mass shell.
We now parametrize the mass shell in coordinates in  which $\sigma$ 
has the standard
form $\sigma^{(0)}$ by $k=(w\cosh\theta,v_{1},w\sinh\theta,v_{2})$ with
$\theta\in\mathbb  R$, $v=(v_1,v_2)\in\mathbb R^2$ and  $w=\sqrt{v^2+m^2}$, such that 
the measure on the mass shell assumes  the form $\frac 1 2\int
d^2v\,d\theta $.  
This may be done without loss of generality, since for any $\sigma\in\Sigma$,
there is an element  $\Lambda$ of the full Lorentz group such that
$\sigma=\Lambda\sigma^{(0)} \Lambda^t$, and thus $k\sigma p=(\Lambda^t k)
\sigma^{(0)}(\Lambda^t p)$. If $\Lambda$ is proper, all $\Lambda^tk_j$,
$j\in U\cup A$ can 
obviously be parametrized by  such coordinates as above, and  if $\Lambda$ is
improper, we
use $k\sigma p=(-\Lambda^t k) \sigma^{(0)}(-\Lambda^t p)$ and parametrize
$-\Lambda^t k_j$,
$j\in U\cup A$ by the above coordinates. In~(\ref{limc}), this amounts to
simply renaming  the arguments. Up to numerical
constants, $\invF{\gamma^{C}(g)}(k_U)$ is therefore given by 
\begin{equation}\label{gamc}
\int d k(\theta,v)_{A}\; 
\exp\big({-i{\textstyle \sum\limits_{s<t} }
J_{st}(w_{s}w_{t}\sinh (\theta_{s}-\theta_{t})+ v_{s}\wedge 
v_{t})}\big) \,,
\end{equation}
where $v_{s}\wedge v_{t}=v_{s,1}v_{t,2}-v_{s,2}v_{t,1}$ and where the indices
$s,t$ are elements of the  index set $U\sqcup A$.
$J_{st}=1$ if the corresponding block of the intersection or 
enclosure matrix is nonzero and $J_{st}=0$ otherwise. 
The integrals over $k_{A}$ are not
absolutely convergent but  oscillatory. To evaluate them, we shift the
integrations over the rapidity  variables $\theta_{A}$ into
the complex plane, $\theta_a+i\eta_a$  such that for $a<a'$, $a,a^\prime \in A$, 
\[
0 < \eta_{a} <  \eta_{a'} < \pi \,.
\]
Using the formulas
\beqa
\sinh (\theta +i\eta)  =  \sinh \theta \cos 
\eta +i\cosh\theta \sin \eta  & \mbox{ and }
&
\cosh (\theta +i\eta)  =  \cosh \theta \cos 
\eta +i\sinh\theta \sin \eta\,,
\eeqa
and setting $\theta_{st}=\theta_s-\theta_t,\eta_{st}=\eta_s-\eta_t$ we may now
replace the integral appearing in (\ref{gamc})  by the following expression:
\begin{equation}\label{gamcmeta} 
\int\!\! d^2 v_{A} 
\!\int \!\! d\theta_{A}\; 
\exp\big(-{i \ts{\sum\limits_{s<t}} J_{st}\left(w_{s}w_{t}\sinh 
\theta_{st}\cos\eta_{st}+v_{s}\wedge v_{t}\right)}
+{\ts{\sum\limits_{s<t}} J_{st}\,w_{s}w_{t} 
\cosh\theta_{st}\sin\eta_{st}}\big) 
\end{equation}
where we put $\eta_u=0$ for $u\in U$.
The integrand decreases fast in the variables $(\theta_{st},J_{st}=1)$, 
since by construction 
\[
\sin\eta_{st}< 0\qquad \mbox{ for all }s<t \in 
U\sqcup A \ .
\]
Since by
definition $C$
is aplanar, all connected components of the contraction have a  nontrivial
enclosure matrix, and we infer that $\exp\big(+{\ts{\sum}
J_{st}\,w_{s}w_{t}  \cosh\theta_{st}\sin\eta_{st}}\big)$ is also fast
decreasing in $\theta_A$: connectedness ensures that all $\theta_A$ appear
at least once and aplanarity ensures that the exponential does not only depend
on the difference variables $\theta_{a a^\prime}$, $a,a^\prime\in A$. Hence,
the integrations over $d\mu_A(k_A)$ are well-defined.
Since furthermore, 
$\check \psi_\omega^{(U)} (-k_U)$ is fast
decreasing in $-\sum_{u\in U} k_u$ (all on the positive mass shell), we may
conclude that  the pure creation parts appearing on the right hand side of
(\ref{limc}) yield well-defined operators in $\mathfrak F$.

An analogous argument shows that the pure annihilation parts ($U'=U$) are
well-defined. In this case, we find 
\[
\invF{\gamma(g)}(k_{U'})=\int d\mu_{A}(k_{A})
e^{-i\langle k_{A},Ik_{A}\rangle -i\langle k_{A},E  k_{U'}\rangle} 
\]
and an analytic continuation $\theta_a+i\eta_a$, $a\in A$, with
$-\pi<\eta_a<\eta_{a'}<0$ for $a<a'$, would yield the desired result since
$\check \psi_\omega^{(U')} (-k_U)$ is fast decreasing in $\sum_{u\in U'} k_u$.
More generally, for contributions with $U'\neq \emptyset$, we have
\[
\invF{\gamma(g)}(k_{U})= \int d\mu_{A}(k_{A}) 
e^{-i\langle k_{A},Ik_{A}\rangle -i\langle k_{A},E (\epsilon_U k_{U})\rangle} \
,
\]
where $\epsilon_Uk_U$ is the tuple $(\epsilon_{u} k_{u})_{u \in U}$,  with
$\epsilon_u=+1$ for $u\in U^\prime$ and $\epsilon_u=-1$ for $u\in U\setminus
U^\prime$. In this case, $\check \psi_\omega (-k_U)$ is fast decreasing
in $-\sum_{u \in U} \epsilon_u k_u $.
We evaluate the expression on a suitable vector in
Fock space to get rid of the annihilation operators and shift the integrations
over the 
rapidity variables $\theta_{A\sqcup U'}$ into the complex plane,
$\theta_s+i\eta_s$ for $s\in A\sqcup U'$, such that $0 < \eta_{s} <  \eta_{t} <
\pi$ for $s<t$, $s,t\in A\sqcup
U'$. Note that in this case, also some of the
arguments of $\check \psi_\omega$ will be analytically continued.  

%%%%%%%%%%%%%%%%%%%

In a similar manner as in the above discussion, we can give meaning to
expressions of the form
\[
\prod_{i=1}^{m} \wickq{\phi^{n_i}(q-x_i)}
\]
which appear in the perturbative  solution of the Yang Feldman equation
(\ref{prodYF}). Here, we may formally write 
\[
\int dx \;\wickq{\phi^{n}(q-x)}\,G(x) \;\stackrel{\rm def}{=}\; 
\wickq{\phi^n_g}(q)
\]
where $g(x_N)=G(x_1)\prod_{j=2}^n \delta (x_1-x_j)$, $N=\{1,\dots,n\}$ with a
suitable test function $G$.
Applying Wick's theorem for quasiplanar Wick products (Theorem~1) to an
expression $\wickq{\phi^n_g}\wickq{\phi^m_f}$, with 
$g(x_N)$ as above and for $M=\{n+1,\dots,n+m\}$,
$f(x_M)=F(x_{n+1})\prod_{j=n+2}^{n+m} \delta (x_{n+1}-x_j)$,
we obtain integrals of the form
\[
\invF{\gamma^C(g\otimes f)}(k_U)=
\int d\mu_{A}(k_{A}) e^{-i(k_{A},Ik_{ A})} \big(\check
G(\ts{\sum\limits_{i \in N}} k_{i}) \check F (\ts{\sum\limits_{j \in M}} k_{j} )
\big)\,\big|_{k_{\alpha( A)}=-k_{ A}} \ , 
\]
where $C\in \mathcal C(N,M)$. Again, we use coordinates $k(\theta,v)$ such that
the twisting is given by $\sigma^{(0)}$ and shift the integration over
$\theta_A$ into the complex plane, $\theta_a+i\eta_a$  such that $0 < \eta_{a}
<  \eta_{a'} < \pi$ for $a<a'$, $a,a^\prime \in A$. We now observe that from the
analytic continuation we obtain the factor 
$\exp\big(+{\ts{\sum_{a<b}}  I_{ab}\,w_{a}w_{b} 
\cosh\theta_{ab}\sin\eta_{ab}}\big)$, $a,b\in  A$, which strongly
decreases in $(\theta_{ab},I_{ab}\neq 0)$. By definition, we have
$\alpha( A_{C'}\cap N)\cap M\neq\emptyset$ for any connected
component $ C^\prime$ of $ C\in \mathcal C(N,M)$. Therefore, 
in any
connected component at least one internal momentum $k_{\tilde A}$ appears {\em
both} in $\check G$ and (with opposite sign) in $\check F$ and
we conclude that the integrand is strongly decreasing in $\theta_A$ such that
the integrals are well-defined.

%%%%%%%%%%%%%%%%%%%

In order to make the above discussion mathematically sound, several details
are missing.  Since they turned out to be quite complicated, we
shall treat them in the forthcoming publication mentioned above, and only name
the necessary steps here. First of all,
we will specify the space of suitable test functions on which the analytic
continuation as performed above is well-defined.  In this test function space,
sequences of functions have to  exist which converge to $\delta$-distributions in an appropriate topology, such that the integrals in
question, evaluated in such sequences, converge to the  expressions discussed
above (in the appropriate topology).  We will moreover show that the Fock space
vectors with wavefunctions from this set of functions form a Lorentz-invariant
stable domain for the quasiplanar Wick products and specify the set of
admissible states $\tilde \omega$ on $\mathcal E$.

%%%%%%%%%%%%%%%%%%%%%%%%

%%%%%%%%%%%%%%%%

\section{Consequences}

In this section we would like to point out some of the consequences of our
analysis. In particular, we comment on the modified dispersion  relation
resulting from the use of quasiplanar Wick products in the perturbative
expansion. While these remarks are not yet conclusive, they provide a hint as to
how the ultraviolet-infrared mixing problem appears in our framework.

The first conclusion we may draw from the previous sections is that 
the divergences discussed here
are not compatible with those arising in a theory on a Euclidean
noncommutative spacetime. To see this, consider the quasiplanar contraction
\begin{picture}(50,9)
\put(5,2){\circle{2}}
\put(15,2){\circle{2}}
\put(25,2){\circle{2}}
\put(35,2){\circle{2}}
\put(45,2){\circle{2}}
\qbezier(5,2)(15,13)(25,2)
\qbezier(15,2)(25,13)(35,2)
\end{picture}. As is well known, on  a Euclidean noncommutative spacetime this
contribution yields a finite result in the limit of coinciding points (i.e. as
a tadpole contribution). This can be understood as follows:  consider a  
test function $f$ in the relative coordinates $x_1-x_3$, $x_2 -x_4$ and in $x_5$ 
which tends to a product of a testfunction $g$ in $x_5$ and
$\delta$-distributions in the relative coordinates. Then on a Euclidean
noncommutative spacetime, we have
\[
\gamma^C_{\rm euc}(f)(x_5)\propto \int d p \int dk
\,\frac{1}{k^2+m^2}\,\frac{1}{p^2+m^2}\,e^{-ikQp}\,(\mathcal F^{-1}_{1,2}
f)(k,p,x_5)\, ,
\]
where $\mathcal F^{-1}_{1,2}$ indicates the inverse Fourier transform with
respect to the first and the second argument. 
Introducing Schwinger parameters and swapping the order of integration, we then 
find 
\[
\int_0^\infty d \alpha d\beta \int dk d \xi \;
e^{-(\alpha+\beta)m^2}\,e^{\alpha k^2}\,e^{-\xi^2/4\beta }\,\pi^2\,\beta^{-2}\, 
(\mathcal F^{-1}_{1} f)(k, Qk -\xi,x_5)\,.
\]
For $Q$ being of maximal rank, this yields a well-defined expression even 
for 
$(\mathcal F^{-1}_1 f)(k, y,x_5)$ tending to $c\,\delta(y)\,g(x_5)$, namely
\[
g(x_5)\int_0^\infty d \alpha d\beta\,
e^{-(\alpha+\beta)m^2}\,\frac{1}{(\alpha \beta + \frac{\lambda_P^4}{4})^2}\,,
\]
where without loss of generality, we have set
$(Qp)^2=\lambda_P^4(p_2^2+p_1^2+p_0^2+p_3^2)$.

In contrast to this, the same contraction is ill-defined on the noncommutative
Minkowski space in the limit of coinciding points. In  order to keep the
calculation simple, we
consider a test function $f$
in the relative variable $x_1-x_3$  with $\check f$ tending to a constant. We
then find $ \gamma^C(f)\propto \int d\mu( p) d\mu (k) \;e^{-ipQk} \, \check
f(p)  \propto \int d\mu(p)\,\check f(p)\,\Delta_+(Qp)$ and  while
$\Delta_+(Qp)$ is a bounded function for $p$ on the positive mass shell, it is
not integrable. To see this, we choose coordinates on the mass shell such
that the twisting is given by $\sigma^{(0)}$ and the argument of the
2-point function is $-\lambda_P^4(p_1^2+p_3^2)$. It follows
that in the limit where $\check f$ tends to a constant, the integration over
$p_2$ diverges logarithmically. This means that results on renormalization
gained in a Euclidean theory may not be directly applied in the Minkowskian
regime.

Furthermore, we would like to emphasize that for commuting time variable, the
quasiplanar Wick products are in general no longer well-defined. To see this,
we first use the fact that an antisymmetric $3\times 3$ matrix has determinant
zero. We can therefore set, without loss of generality, $Q p=\lambda_P^2
(0,p_3,0,-p_1)$. Already the simplest aplanar contraction, \begin{picture}(27,10)
\put(2,1){\circle{2}}
\put(12,1){\circle{2}}
\put(22,1){\circle{2}}
\qbezier(2,1)(12,14)(22,1)
\end{picture} becomes ill-defined in the limit of coinciding points, 
since (contrary to the case where $Q$ is nondegenerate)
it contains the ill-defined integral
$\int\! d \mu(p) \,\exp\big({-i\lambda_P^2(p_3k_1 -p_1k_3)}\big)$. Since the
contraction still violates the locality condition, it follows that such a
theory is not renormalizable by local counterterms\footnote{The ill-definedness
of the contraction may also be understood by the fact that for the $Q$ under
consideration,
$\Delta_+(x+Qk)$ cannot be multiplied (as a distribution) with
$\delta(x_0)\delta(x_2)$.}. See also~\cite{ggbrr}.

The application of quasiplanar Wick products in the framework of the
Yang-Feldman equation is straightforward. In the rules spelled out explicitly 
in~\cite{bahnsdiss} for ordinary Wick products, one only has to replace the
Wick products by quasiplanar Wick products. From preliminary calculations we
have performed at lower orders of the perturbative expansion, it is reasonable
to hope  that quasiplanar counterterms suffice as counterterms to render the
theory ultraviolet finite. However, if we employ the quasiplanar Wick products
and thus refrain from subtracting nonlocal counterterms, we encounter a serious
modification of the dispersion relation. Similar discussions in the context of
 space-space-noncommutativity, which are not founded on the general construction of
quasiplanar  Wick products, may be found in~\cite{sib,alvgenprop}.

Let us assume that all ultraviolet divergent terms can be absorbed in
quasiplanar (thus local) counterterms, leading in  particular to a finite mass
$m$ in the renormalized field equation,
\[
(\square_q +
m^2)\,\phi(q)\;=\;-g\,\phi^{n-1}(q)+(\underbrace{m^2-m_0^2}_{\displaystyle
=\delta m^2})\,\phi(q) +\cdots 
\] 
where $m_0$ is the bare mass and the dots indicate the remaining counterterms
(starting with order $g^2$).  If we now insert the renormalized field  as a
formal power series in $g$, we find at lowest order, for $n=4$,
\begin{equation}\label{phidis}
(\square + m^2)(\phi_0(q)+\dots)\;=\;
-g\,\phi_0^3(q)+\delta m^2_1\,\phi_0(q) +\cdots
\end{equation}
Now according to our programme, 
\[
-g\,\phi_0^3(q)+\delta m^2_1\,\phi_0(q)\;=\;
-g\,\wickq{\phi_0^3(q)} \;=\;
-g\,\wickn{\phi_0^3(q)}-g \begin{picture}(30,10)(-5,0)
\put(0,2){\circle{2}}
\put(10,2){\circle{2}}
\put(20,2){\circle{2}}
\qbezier(0,2)(10,15)(20,2)
\end{picture}\ ,
\] 
such that taking the expectation value $\langle 0|\;\cdot\;|p\rangle$ on both
sides of equation~(\ref{phidis}), we find a modification of the ordinary
dispersion relation of the following form,
\[
-p^2+m^2=-g\,\Delta_+(Qp)+\dots\ ,
\]
where $\Delta_+$ is the 2-point function at mass $m$.

Allowing for additional counterterms, $\alpha$ and $\beta p^2$, we thus find at
this order
\[
p^2-m^2-g\,(\Delta_+(Qp)+\alpha_1 +\beta_1 p^2) = 0\ .
\]

We now choose the fixed value $\sigma^{(0)}$ for $Q$. 
Then the transversal velocity $v_\perp=(v_1,v_3)$ is
\[
v_\perp=\nabla_{p_\perp}\; p_0
=
\frac{p_\perp}{p_0} \;\;
\frac{1+ 
\,\frac{g}{1-g\beta_1 }
\,\,\eta(p)}
{1- \,\frac{g}{1-g\beta_1}\,
\,\,\eta(p)}\,,
\]
where $
\eta(p)
=(2p_1)^{-1}\,\partial_{p_1}\,\Delta_+(\sigma^{(0)}p)
=(2p_3)^{-1}\,\partial_{p_3}\,\Delta_+(\sigma^{(0)}p)
=-\frac{m^2\,K_2(\lambda_P^2 m\,{\sqrt{{{p_0}}^2  - {{p_2}}^2 + 
{{p_\perp}}^2}})  }{8\,{\pi }^2\,
\left( {{p_0}}^2  - {{p_2}}^2 + {{p_\perp}}^2\right) }
$
and therefore depends only on $(\sigma^{(0)}p)^2$. Now assume that $p$ is on
the physical mass shell, $p^2=M^2$, where $M$ is allowed to be different from 
$m$ (though the latter is finite), then 
\[
\eta(p)\,\big|_{p^2=M^2}=-\frac{
m^2\,K_2(
\lambda_P^2 m\,M\,\sqrt{1+\tsf{2\,p_\perp^2}{M^2}})  }{8\,{\pi }^2\,
M^2\,\left(1+\tsf{2\,p_\perp^2}{M^2}\right) }\,.
\]
If the masses $m$ and $M$ are both assumed to be 
of the order of the Planck mass, the factor
$\frac{1+  \,\frac{g}{1-g\beta_1 } \,\,\eta(p)} {1-
\,\frac{g}{1-g\beta_1}\, \,\,\eta(p)}$ as a function of the transversal
component  $p_\perp$ is of the following form:
\[
\epsfig{figure=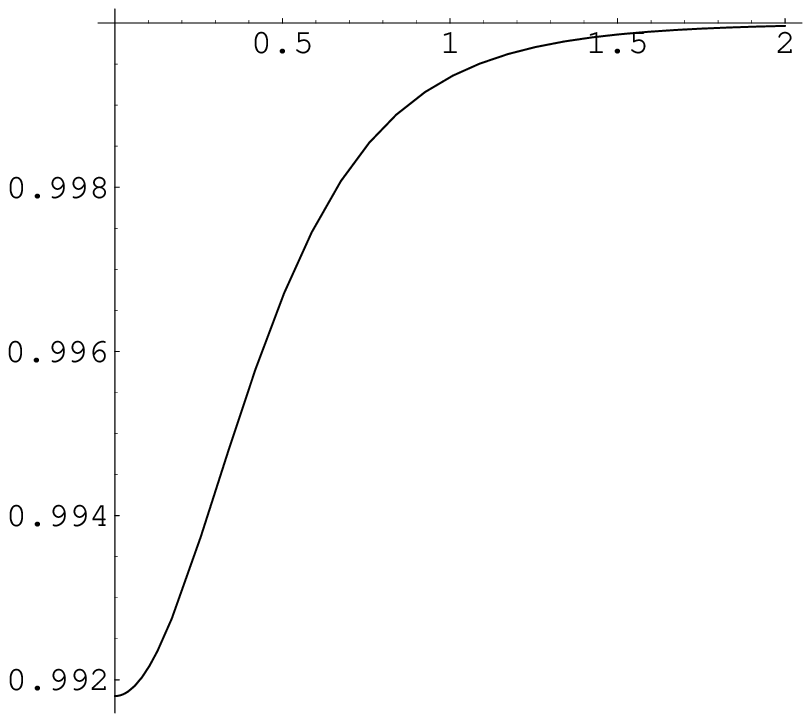, height=5cm, width=7cm}
\mbox{\small \begin{tabular}{l}$m=M=1$, $\beta_1=0$,  $g=1/5$,
 $\lambda_P=1$\\
plotted with {\tt Mathematica}.\end{tabular}}
\]
Surprisingly, the maximal deviation does not occur at
high momenta but at $p_\perp=0$. In the above numerical setting, this point of 
maximal deviation is of the order of $1 \% 
$,
\[
(1+g\tilde\eta(m))/(1-g\tilde\eta(m))\Big|_{m^2=1}\simeq 0.99\,,\qquad
\tilde\eta(m)
\stackrel{\rm def}{=}
\eta(p)\,\big|_{\substack{p_\perp=0\\p^2=m^2}}=-(8\pi^2)^{-1}\,
K_2(\lambda_P^2 m^2)\ .
\]
Using  smaller masses $m=M< m_P$, the deviation becomes even
larger, as we can see in the following plot, where 
$\frac{1+g\,\tilde\eta(m)}{1-g\,\tilde\eta(m)}$ (i.e. the maximal deviation from
$1$)  is plotted as a function of the mass $m$, ranging from 0 to 1:
\[
\epsfig{figure=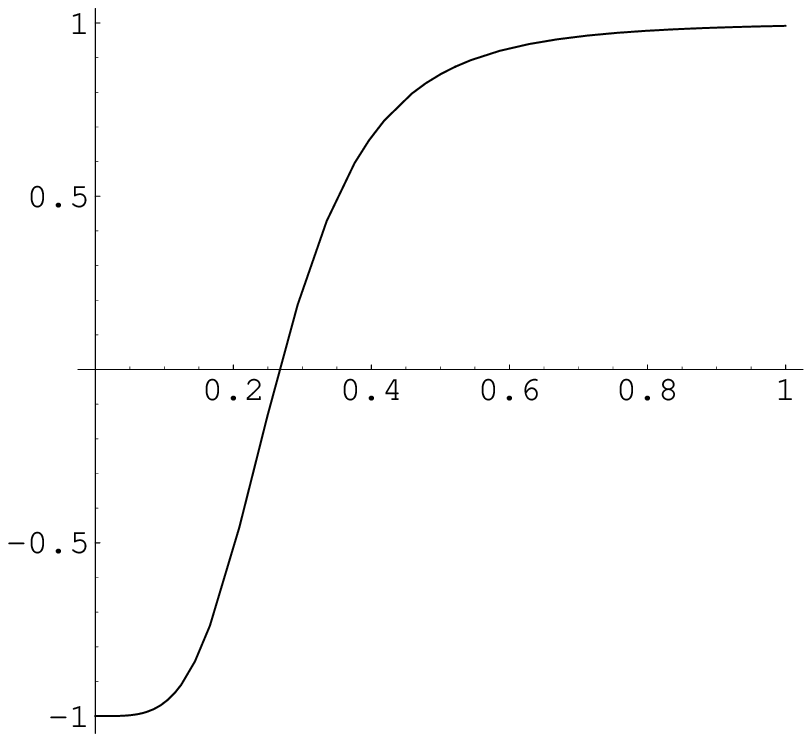, height=5cm, width=7cm}
\mbox{\small \begin{tabular}{l}
$M=m$, 
$\beta_1=0$,  $g=1/5$, $\lambda_P=1$\\
plotted with {\tt Mathematica}.\end{tabular}}
\]
We see that the group velocity may even become negative.
Integrating over, say, $\Sigma_1$ would not improve the situation: since the
scale $\lambda_P$ remains fixed,  the behaviour sketched above would
qualitatively remain the same.

If we take into account that $m$ and $M$ may differ from one another, it
is possible to allow for small physical masses $M$ while taking $m$ to be very
large. To see this, observe that at $p_\perp=0$ and $p^2=M^2$,
\[
\eta(p)=-\frac{1 }{8\,{\pi }^2}
\,\frac{m^2}{M^2} \,K_2\big(
      \lambda_P^2  \,M^2\,\frac m M \big)\,,
\]
and since $\alpha^2 K_2(\beta\,\alpha)\longrightarrow  0$ for $\alpha$ large
enough, it is possible to make the deviation arbitrarily small even for small
masses by choosing $m$ large enough. It remains to be investigated whether this
scheme can be applied consistently to all orders, but in any case it would be a
``finetuning'' procedure which does not seem to be very natural.

However small, the modification of the dispersion relation has serious
consequences. In ordinary local quantum field theory, the Hilbert space of the
asymptotic fields is the Fock space of the free fields with fixed (constant)
mass. The above analysis shows that this cannot be true for the asymptotic
fields in the framework considered here, since their mass will in general
depend on the momentum. In a realistic model such as quantum electrodynamics,
the modified dispersion relation could provide predictions which by comparison
with experiment might seriously restrict the scale of noncommutativity. 
In
the above, this scale was taken to be of the order of the Planck length.
The effect being larger for a smaller parameter $\lambda_P$ (i.e. for a higher
energy), it is not impossible that in a realistic model such as quantum
electrodynamics, where phenomenological calculations so far have provided lower
bounds for the energy scale of noncommutativity, an {\em upper} bound for the
energy scale could be derived in this way -- depending on how questions
concerning renormalization  can be solved.

%%%%%%%%%%%%%%%%%%%%%%%%%%%%%%%%%%%%%%%%%%
\begin{appendix}

\section{Examples}

In the following examples, quasiplanar Wick products are symbolized by
boxes, and contractions by connecting
lines as explained on page~\pageref{graphs}.

\subsection{Formula (\ref{qpWick})}\label{exqpWick}

\beqa
\begin{picture}(50,15)
\put(0,2){\framebox{\parbox[t]{42.5pt}{$\phantom{\dots}$}}}
\put(5,2){\circle{2}}
\put(15,2){\circle{2}}
\put(25,2){\circle{2}}
\put(35,2){\circle{2}}
\put(45,2){\circle{2}}
\end{picture}& = &
\begin{picture}(50,15)
\put(5,2){\circle{2}}
\put(15,2){\circle{2}}
\put(25,2){\circle{2}}
\put(35,2){\circle{2}}
\put(45,2){\circle{2}}
\end{picture}\ -\
\begin{picture}(50,15)
\put(5,2){\circle{2}}
\put(15,2){\circle{2}}
\put(25,2){\circle{2}}
\put(35,2){\circle{2}}
\put(45,2){\circle{2}}
\qbezier(35,2)(40,11)(45,2)
\end{picture}\ -\
\begin{picture}(50,15)
\put(5,2){\circle{2}}
\put(15,2){\circle{2}}
\put(25,2){\circle{2}}
\put(35,2){\circle{2}}
\put(45,2){\circle{2}}
\qbezier(25,2)(30,11)(35,2)
\end{picture}\ -\
\begin{picture}(50,15)
\put(5,2){\circle{2}}
\put(15,2){\circle{2}}
\put(25,2){\circle{2}}
\put(35,2){\circle{2}}
\put(45,2){\circle{2}}
\qbezier(15,2)(20,11)(25,2)
\end{picture}\ -\
\begin{picture}(50,15)
\put(5,2){\circle{2}}
\put(15,2){\circle{2}}
\put(25,2){\circle{2}}
\put(35,2){\circle{2}}
\put(45,2){\circle{2}}
\qbezier(5,2)(10,11)(15,2)
\end{picture}\\
&&-\
\begin{picture}(50,15)
\put(5,2){\circle{2}}
\put(15,2){\circle{2}}
\put(25,2){\circle{2}}
\put(35,2){\circle{2}}
\put(45,2){\circle{2}}
\qbezier(15,2)(25,17)(35,2)
\qbezier(25,2)(35,17)(45,2)
\end{picture}\ -\
\begin{picture}(50,15)
\put(5,2){\circle{2}}
\put(15,2){\circle{2}}
\put(25,2){\circle{2}}
\put(35,2){\circle{2}}
\put(45,2){\circle{2}}
\qbezier(5,2)(15,17)(25,2)
\qbezier(15,2)(25,17)(35,2)
\end{picture}\ + \
\begin{picture}(50,15)
\put(5,2){\circle{2}}
\put(15,2){\circle{2}}
\put(25,2){\circle{2}}
\put(35,2){\circle{2}}
\put(45,2){\circle{2}}
\qbezier(15,2)(20,11)(25,2)
\qbezier(35,2)(40,11)(45,2)
\end{picture}\ + \
\begin{picture}(50,15)
\put(5,2){\circle{2}}
\put(15,2){\circle{2}}
\put(25,2){\circle{2}}
\put(35,2){\circle{2}}
\put(45,2){\circle{2}}
\qbezier(5,2)(10,11)(15,2)
\qbezier(35,2)(40,11)(45,2)
\end{picture}\ +\
\begin{picture}(50,15)
\put(5,2){\circle{2}}
\put(15,2){\circle{2}}
\put(25,2){\circle{2}}
\put(35,2){\circle{2}}
\put(45,2){\circle{2}}
\qbezier(5,2)(10,11)(15,2)
\qbezier(25,2)(30,11)(35,2)
\end{picture}
\eeqa

%Compare this with the ordinary Wick product as given in formula (\ref{ordWick}), 

%%%%%%%%%%%%%%%%%

\subsection{Formula (\ref{qprecursion})}\label{exqprecursion}

\beqa
\begin{picture}(50,15)
\put(0,2){\framebox{\parbox[t]{42.5pt}{$\phantom{\dots}$}}}
\put(5,2){\circle{2}}
\put(15,2){\circle{2}}
\put(25,2){\circle{2}}
\put(35,2){\circle{2}}
\put(45,2){\circle{2}}
\end{picture}& = &
\begin{picture}(50,15)
\put(10,2){\framebox{\parbox[t]{32.5pt}{$\phantom{\dots}$}}}
\put(5,2){\circle{2}}
\put(15,2){\circle{2}}
\put(25,2){\circle{2}}
\put(35,2){\circle{2}}
\put(45,2){\circle{2}}
\end{picture}\ -\
\begin{picture}(50,15)
\put(20,2){\framebox{\parbox[t]{22.5pt}{$\phantom{\dots}$}}}
\put(5,2){\circle{2}}
\put(15,2){\circle{2}}
\put(25,2){\circle{2}}
\put(35,2){\circle{2}}
\put(45,2){\circle{2}}
\qbezier(5,2)(10,11)(15,2)
\end{picture}\ -\
\begin{picture}(50,15)
\put(5,2){\circle{2}}
\put(15,2){\circle{2}}
\put(25,2){\circle{2}}
\put(35,2){\circle{2}}
\put(45,2){\circle{2}}
\qbezier(5,2)(15,17)(25,2)
\qbezier(15,2)(25,17)(35,2)
\end{picture}
\\
\begin{picture}(40,15)
\put(0,2){\framebox{\parbox[t]{32.5pt}{$\phantom{\dots}$}}}
\put(5,2){\circle{2}}
\put(15,2){\circle{2}}
\put(25,2){\circle{2}}
\put(35,2){\circle{2}}
\end{picture}& = &
\begin{picture}(40,15)
\put(10,2){\framebox{\parbox[t]{22.5pt}{$\phantom{\dots}$}}}
\put(5,2){\circle{2}}
\put(15,2){\circle{2}}
\put(25,2){\circle{2}}
\put(35,2){\circle{2}}
\end{picture}\ - \
\begin{picture}(40,15)
\put(20,2){\framebox{\parbox[t]{12.5pt}{$\phantom{\dots}$}}}
\put(5,2){\circle{2}}
\put(15,2){\circle{2}}
\put(25,2){\circle{2}}
\put(35,2){\circle{2}}
\qbezier(5,2)(10,11)(15,2)
\end{picture}\ - \
\begin{picture}(40,15)
\put(5,2){\circle{2}}
\put(15,2){\circle{2}}
\put(25,2){\circle{2}}
\put(35,2){\circle{2}}
\qbezier(5,2)(15,17)(25,2)
\qbezier(15,2)(25,17)(35,2)
\end{picture}
\\
\begin{picture}(30,15)
\put(0,2){\framebox{\parbox[t]{22.5pt}{$\phantom{\dots}$}}}
\put(5,2){\circle{2}}
\put(15,2){\circle{2}}
\put(25,2){\circle{2}}
\end{picture} &=&
\begin{picture}(30,15)
\put(10,2){\framebox{\parbox[t]{12.5pt}{$\phantom{\dots}$}}}
\put(5,2){\circle{2}}
\put(15,2){\circle{2}}
\put(25,2){\circle{2}}
\end{picture}\ -\
\begin{picture}(30,15)
\put(5,2){\circle{2}}
\put(15,2){\circle{2}}
\put(25,2){\circle{2}}
\qbezier(5,2)(10,11)(15,2)
\end{picture}
\\
\begin{picture}(20,15)
\put(0,2){\framebox{\parbox[t]{12.5pt}{$\phantom{\dots}$}}}
\put(5,2){\circle{2}}
\put(15,2){\circle{2}}
\end{picture} &=&
\begin{picture}(20,15)
\put(5,2){\circle{2}}
\put(15,2){\circle{2}}
\end{picture}\ -\
\begin{picture}(20,15)
\put(5,2){\circle{2}}
\put(15,2){\circle{2}}
\qbezier(5,2)(10,11)(15,2)
\end{picture}
\eeqa

\subsection{Formula (\ref{qpWickthm})}\label{wickex}

In what follows, the underscore symbolizes quasiplanar Wick ordering of fields
which are not direct neighbours. For instance, for  the contraction
$C\in\mathcal C((1,\dots,4)\sqcup (5,\dots,8))$ with $U_C=(1,2,3,8)$,
$A_C=(4,5)$, $\alpha(4)=6$ and $\alpha(5)=7$, we write
\[
\wickq{\phi^{|U|}_{\gamma^C(f)}}
\;\;=\;\;
\begin{picture}(90,15)
\put(0,2){\framebox{\parbox[t]{22.5pt}{$\phantom{\dots}$}}}
\put(20,0.4){\underline{\parbox[b]{50pt}{$\phantom{\dots}$}}}
\put(70,2){\framebox{\parbox[t]{4.5pt}{$\phantom{\dots}$}}}
\qbezier(35,2)(45,17)(55,2)
\qbezier(45,2)(55,17)(65,2)
\put(5,2){\circle{2}}
\put(15,2){\circle{2}}
\put(65,2){\circle{2}}
\put(75,2){\circle{2}}
\put(25,2){\circle{2}}
\put(35,2){\circle{2}}
\put(45,2){\circle{2}}
\put(55,2){\circle{2}}
\put(40,-4){\line(0,0){6}}
\end{picture}\ ,
\]
where the small vertical line serves to separate the sets 
$(1,\dots,4)$ and $(5,\dots,8)$ from one another.

\vspace{2ex}
{\bf Example 1:}
\vspace{1ex}

\begin{picture}(90,10)
\put(0,2){\framebox{\parbox[t]{32.5pt}{$\phantom{\dots}$}}}
\put(5,2){\circle{2}}
\put(15,2){\circle{2}}
\put(25,2){\circle{2}}
\put(35,2){\circle{2}}
\put(40,2){\framebox{\parbox[t]{32.5pt}{$\phantom{\dots}$}}}
\put(45,2){\circle{2}}
\put(55,2){\circle{2}}
\put(65,2){\circle{2}}
\put(75,2){\circle{2}}
\end{picture}
$=\quad$
\begin{picture}(90,10)
\put(0,2){\framebox{\parbox[t]{75pt}{$\phantom{\dots}$}}}
\put(5,2){\circle{2}}
\put(15,2){\circle{2}}
\put(65,2){\circle{2}}
\put(75,2){\circle{2}}
\put(25,2){\circle{2}}
\put(35,2){\circle{2}}
\put(45,2){\circle{2}}
\put(55,2){\circle{2}}
\put(40,-4){\line(0,0){6}}
\end{picture}
$+\quad$
\begin{picture}(90,10)
\put(0,2){\framebox{\parbox[t]{22.5pt}{$\phantom{\dots}$}}}
\put(20,0.5){\underline{\parbox[b]{50pt}{$\phantom{\dots}$}}}
\put(50,2){\framebox{\parbox[t]{22.5pt}{$\phantom{\dots}$}}}
\qbezier(35,2)(40,11)(45,2)
\put(5,2){\circle{2}}
\put(15,2){\circle{2}}
\put(65,2){\circle{2}}
\put(75,2){\circle{2}}
\put(25,2){\circle{2}}
\put(35,2){\circle{2}}
\put(45,2){\circle{2}}
\put(55,2){\circle{2}}
\put(40,-4){\line(0,0){6}}
\end{picture}
$+\quad$
\begin{picture}(90,10)
\put(0,2){\framebox{\parbox[t]{4.5pt}{$\phantom{\dots}$}}}
\put(10,0.5){\underline{\parbox[b]{50pt}{$\phantom{\dots}$}}}
\put(50,2){\framebox{\parbox[t]{22.5pt}{$\phantom{\dots}$}}}
\qbezier(15,2)(25,17)(35,2)
\qbezier(25,2)(35,17)(45,2)
\put(5,2){\circle{2}}
\put(15,2){\circle{2}}
\put(65,2){\circle{2}}
\put(75,2){\circle{2}}
\put(25,2){\circle{2}}
\put(35,2){\circle{2}}
\put(45,2){\circle{2}}
\put(55,2){\circle{2}}
\put(40,-4){\line(0,0){6}}
\end{picture}

\begin{picture}(90,30)
\end{picture}
$+\quad$
\begin{picture}(90,30)
\put(0,2){\framebox{\parbox[t]{22.5pt}{$\phantom{\dots}$}}}
\put(20,0.5){\underline{\parbox[b]{50pt}{$\phantom{\dots}$}}}
\put(70,2){\framebox{\parbox[t]{4.5pt}{$\phantom{\dots}$}}}
\qbezier(35,2)(45,17)(55,2)
\qbezier(45,2)(55,17)(65,2)
\put(5,2){\circle{2}}
\put(15,2){\circle{2}}
\put(65,2){\circle{2}}
\put(75,2){\circle{2}}
\put(25,2){\circle{2}}
\put(35,2){\circle{2}}
\put(45,2){\circle{2}}
\put(55,2){\circle{2}}
\put(40,-4){\line(0,0){6}}
\end{picture}
$+\quad$
\begin{picture}(90,30)
\put(0,2){\framebox{\parbox[t]{15pt}{$\phantom{\dots}$}}}
\put(20,0.5){\underline{\parbox[b]{50pt}{$\phantom{\dots}$}}}
\put(60,2){\framebox{\parbox[t]{15pt}{$\phantom{\dots}$}}}
\qbezier(25,2)(35,17)(45,2)
\qbezier(35,2)(45,17)(55,2)
\put(5,2){\circle{2}}
\put(15,2){\circle{2}}
\put(65,2){\circle{2}}
\put(75,2){\circle{2}}
\put(25,2){\circle{2}}
\put(35,2){\circle{2}}
\put(45,2){\circle{2}}
\put(55,2){\circle{2}}
\put(40,-4){\line(0,0){6}}
\end{picture}
$+\quad$
\begin{picture}(90,30)
\put(0,2){\framebox{\parbox[t]{15pt}{$\phantom{\dots}$}}}
\put(20,0.5){\underline{\parbox[b]{50pt}{$\phantom{\dots}$}}}
\put(60,2){\framebox{\parbox[t]{15pt}{$\phantom{\dots}$}}}
\qbezier(25,2)(40,19)(55,2)
\qbezier(35,2)(40,11)(45,2)
\put(5,2){\circle{2}}
\put(15,2){\circle{2}}
\put(65,2){\circle{2}}
\put(75,2){\circle{2}}
\put(25,2){\circle{2}}
\put(35,2){\circle{2}}
\put(45,2){\circle{2}}
\put(55,2){\circle{2}}
\put(40,-4){\line(0,0){6}}
\end{picture}

%%%
%%%

\begin{picture}(90,30)
\end{picture}
$+\quad$
\begin{picture}(90,30)
\put(0,2){\framebox{\parbox[t]{15pt}{$\phantom{\dots}$}}}
\qbezier(25,2)(35,19)(65,2)
\qbezier(35,2)(40,11)(45,2)
\qbezier(55,2)(65,19)(75,2)
\put(5,2){\circle{2}}
\put(15,2){\circle{2}}
\put(25,2){\circle{2}}
\put(35,2){\circle{2}}
\put(45,2){\circle{2}}
\put(55,2){\circle{2}}
\put(65,2){\circle{2}}
\put(75,2){\circle{2}}
\put(40,-4){\line(0,0){6}}
\end{picture}
$+\quad$
\begin{picture}(90,30)
\put(0,2){\framebox{\parbox[t]{15pt}{$\phantom{\dots}$}}}
\qbezier(25,2)(50,19)(75,2)
\qbezier(35,2)(50,13)(55,2)
\qbezier(45,2)(50,13)(65,2)
\put(5,2){\circle{2}}
\put(15,2){\circle{2}}
\put(25,2){\circle{2}}
\put(35,2){\circle{2}}
\put(45,2){\circle{2}}
\put(55,2){\circle{2}}
\put(65,2){\circle{2}}
\put(75,2){\circle{2}}
\put(40,-4){\line(0,0){6}}
\end{picture}
$+\quad$
\begin{picture}(90,30)
\put(0,2){\framebox{\parbox[t]{15pt}{$\phantom{\dots}$}}}
\put(38,7){$C_6$}
\put(5,2){\circle{2}}
\put(15,2){\circle{2}}
\put(25,2){\circle{2}}
\put(35,2){\circle{2}}
\put(45,2){\circle{2}}
\put(55,2){\circle{2}}
\put(65,2){\circle{2}}
\put(75,2){\circle{2}}
\put(40,-4){\line(0,0){6}}
\end{picture}

\begin{picture}(90,30)
\end{picture}
$+\quad$
\begin{picture}(90,30)
\put(0,2){\framebox{\parbox[t]{4.5pt}{$\phantom{\dots}$}}}
\put(0,0.3){\underline{\parbox[b]{70pt}{$\phantom{\dots}$}}}
\put(70,2){\framebox{\parbox[t]{4.5pt}{$\phantom{\dots}$}}}
\qbezier(15,2)(40,19)(65,2)
\qbezier(25,2)(40,13)(55,2)
\qbezier(35,2)(40,7)(45,2)
\put(5,2){\circle{2}}
\put(15,2){\circle{2}}
\put(25,2){\circle{2}}
\put(35,2){\circle{2}}
\put(45,2){\circle{2}}
\put(55,2){\circle{2}}
\put(65,2){\circle{2}}
\put(75,2){\circle{2}}
\put(40,-4){\line(0,0){6}}
\end{picture}
$+\quad$
\begin{picture}(90,30)
\put(0,2){\framebox{\parbox[t]{4.5pt}{$\phantom{\dots}$}}}
\put(0,0.3){\underline{\parbox[b]{70pt}{$\phantom{\dots}$}}}
\put(70,2){\framebox{\parbox[t]{4.5pt}{$\phantom{\dots}$}}}
\qbezier(15,2)(40,19)(55,2)
\qbezier(25,2)(50,19)(65,2)
\qbezier(35,2)(40,11)(45,2)
\put(5,2){\circle{2}}
\put(15,2){\circle{2}}
\put(25,2){\circle{2}}
\put(35,2){\circle{2}}
\put(45,2){\circle{2}}
\put(55,2){\circle{2}}
\put(65,2){\circle{2}}
\put(75,2){\circle{2}}
\put(40,-4){\line(0,0){6}}
\end{picture}
$+\quad$
\begin{picture}(90,30)
\put(0,2){\framebox{\parbox[t]{4.5pt}{$\phantom{\dots}$}}}
\put(0,0.3){\underline{\parbox[b]{70pt}{$\phantom{\dots}$}}}
\put(70,2){\framebox{\parbox[t]{4.5pt}{$\phantom{\dots}$}}}
\qbezier(15,2)(40,19)(65,2)
\qbezier(25,2)(43,11)(45,2)
\qbezier(35,2)(43,11)(55,2)
\put(5,2){\circle{2}}
\put(15,2){\circle{2}}
\put(25,2){\circle{2}}
\put(35,2){\circle{2}}
\put(45,2){\circle{2}}
\put(55,2){\circle{2}}
\put(65,2){\circle{2}}
\put(75,2){\circle{2}}
\put(40,-4){\line(0,0){6}}
\end{picture}

\begin{picture}(90,30)
\end{picture}
$+\quad$
\begin{picture}(90,30)
\put(0,2){\framebox{\parbox[t]{4.5pt}{$\phantom{\dots}$}}}
\put(0,0.3){\underline{\parbox[b]{70pt}{$\phantom{\dots}$}}}
\put(70,2){\framebox{\parbox[t]{4.5pt}{$\phantom{\dots}$}}}
\put(38,7){$C_6$}
\put(5,2){\circle{2}}
\put(15,2){\circle{2}}
\put(25,2){\circle{2}}
\put(35,2){\circle{2}}
\put(45,2){\circle{2}}
\put(55,2){\circle{2}}
\put(65,2){\circle{2}}
\put(75,2){\circle{2}}
\put(40,-4){\line(0,0){6}}
\end{picture}
$+\quad$
\begin{picture}(90,30)
\put(60,2){\framebox{\parbox[t]{15pt}{$\phantom{\dots}$}}}
\qbezier(5,2)(15,19)(25,2)
\qbezier(35,2)(40,11)(45,2)
\qbezier(15,2)(40,19)(55,2)
\put(5,2){\circle{2}}
\put(15,2){\circle{2}}
\put(25,2){\circle{2}}
\put(35,2){\circle{2}}
\put(45,2){\circle{2}}
\put(55,2){\circle{2}}
\put(65,2){\circle{2}}
\put(75,2){\circle{2}}
\put(40,-4){\line(0,0){6}}
\end{picture}
$+\quad$
\begin{picture}(90,30)
\put(60,2){\framebox{\parbox[t]{15pt}{$\phantom{\dots}$}}}
\qbezier(5,2)(30,19)(55,2)
\qbezier(15,2)(30,13)(35,2)
\qbezier(25,2)(30,13)(45,2)
\put(5,2){\circle{2}}
\put(15,2){\circle{2}}
\put(25,2){\circle{2}}
\put(35,2){\circle{2}}
\put(45,2){\circle{2}}
\put(55,2){\circle{2}}
\put(65,2){\circle{2}}
\put(75,2){\circle{2}}
\put(40,-4){\line(0,0){6}}
\end{picture}

\begin{picture}(90,30)
\end{picture}
$+\;\;\;$
\begin{picture}(90,30)
\put(60,2){\framebox{\parbox[t]{15pt}{$\phantom{\dots}$}}}
\put(33,7){$C_6$}
\put(5,2){\circle{2}}
\put(15,2){\circle{2}}
\put(25,2){\circle{2}}
\put(35,2){\circle{2}}
\put(45,2){\circle{2}}
\put(55,2){\circle{2}}
\put(65,2){\circle{2}}
\put(75,2){\circle{2}}
\put(40,-4){\line(0,0){6}}
\end{picture}
$+\quad$
\begin{picture}(90,30)
\put(35,7){$\Delta_{4|4}$}
\put(5,2){\circle{2}}
\put(15,2){\circle{2}}
\put(65,2){\circle{2}}
\put(75,2){\circle{2}}
\put(25,2){\circle{2}}
\put(35,2){\circle{2}}
\put(45,2){\circle{2}}
\put(55,2){\circle{2}}
\put(40,-4){\line(0,0){6}}
\end{picture}

\vspace{1ex}
where 
\[
C_{6}=\sum_{\substack{C\in \mathcal C(N)
\\C\; {\rm connected}\\U_C=\emptyset}}\!\!\! \gamma^C(f)
\;\;=\;\;
\begin{picture}(65,10)
\qbezier(5,2)(15,19)(25,2)
\qbezier(15,2)(30,22)(45,2)
\qbezier(35,2)(45,19)(55,2)
\put(5,2){\circle{2}}
\put(15,2){\circle{2}}
\put(25,2){\circle{2}}
\put(35,2){\circle{2}}
\put(45,2){\circle{2}}
\put(55,2){\circle{2}}
\end{picture}
+
\begin{picture}(65,10)
\qbezier(5,2)(25,19)(45,2)
\qbezier(15,2)(25,11)(35,2)
\qbezier(25,2)(40,19)(55,2)
\put(5,2){\circle{2}}
\put(15,2){\circle{2}}
\put(25,2){\circle{2}}
\put(35,2){\circle{2}}
\put(45,2){\circle{2}}
\put(55,2){\circle{2}}
\end{picture}
+\begin{picture}(65,10)
\qbezier(5,2)(20,19)(35,2)
\qbezier(15,2)(35,24)(55,2)
\qbezier(25,2)(35,17)(45,2)
\put(5,2){\circle{2}}
\put(15,2){\circle{2}}
\put(25,2){\circle{2}}
\put(35,2){\circle{2}}
\put(45,2){\circle{2}}
\put(55,2){\circle{2}}
\end{picture}
+
\begin{picture}(65,10)
\qbezier(5,2)(15,19)(35,2)
\qbezier(15,2)(25,22)(45,2)
\qbezier(25,2)(40,17)(55,2)
\put(5,2){\circle{2}}
\put(15,2){\circle{2}}
\put(25,2){\circle{2}}
\put(35,2){\circle{2}}
\put(45,2){\circle{2}}
\put(55,2){\circle{2}}
\end{picture}
\]
with $N=(1,\dots,6)$, and where 
\[
\Delta_{4|4}\hspace{1ex}
=\sum_{\substack{C\in \mathcal C(N\sqcup M) \\ U_C=\emptyset}} \gamma^C(f) 
\hspace{1ex}=\sum_{\substack{C\in \mathcal C(N\sqcup M) \\ 
\substack{A_C=(1,2,3,4)}} }
\!\!\!\gamma^C(f)
\hspace{1ex}+\hspace{1ex}\sum_{i=1}^9
\sum_{C_i\in \mathcal C(N\sqcup M)}\hspace{-1.7ex}\gamma^{C_i}(f)
\]
with $N=(1,2,3,4)$, $M=(5,6,7,8)$, and with the pairs $(A_i,\alpha_i)$
of the contractions $C_i$ determined by 
\[
\begin{array}{lcl}
A_1=(1,2,4,5)\,,\; \alpha_1(1)=3, \alpha_1(5)=7&& 
A_2=(1,2,4,5)\,,\;\alpha_2(1)=3, \alpha_2(5)=8\\
A_3=(1,2,4,6)\,,\;\alpha_3(1)=3, \alpha_3(6)=8&&
A_4=(1,2,3,5)\,,\; \alpha_4(2)=4, \alpha_4(5)=7\\
A_5=(1,2,3,5)\,,\; \alpha_5(2)=4, \alpha_5(5)=8&&
A_6=(1,2,3,6)\,,\;\alpha_6(2)=4, \alpha_6(6)=8\\
A_7=(1,2,3,5)\,,\;\alpha_7(1)=4, \alpha_7(5)=7&&
A_8=(1,2,3,5)\,,\;\alpha_8(1)=4, \alpha_8(5)=8\\
A_9=(1,2,3,6)\,,\;\alpha_9(1)=4, \alpha_9(6)=8&&
\end{array}
\]
such that for instance,
\[
\sum_{C_1\in \mathcal C(N\sqcup M) }
\hspace{-2ex}\gamma^{C_1}(f)
\hspace{2ex}=
\hspace{2ex}\begin{picture}(80,10)
\put(5,2){\circle{2}}
\put(15,2){\circle{2}}
\put(65,2){\circle{2}}
\put(75,2){\circle{2}}
\put(25,2){\circle{2}}
\put(35,2){\circle{2}}
\put(45,2){\circle{2}}
\put(55,2){\circle{2}}
\put(40,-4){\line(0,0){6}}
\qbezier(5,2)(15,17)(25,2)
\qbezier(45,2)(55,17)(65,2)
\qbezier(15,2)(40,22)(55,2)
\qbezier(35,2)(55,24)(75,2)
\end{picture}
\hspace{2ex}+
\hspace{2ex}
\begin{picture}(80,10)
\put(5,2){\circle{2}}
\put(15,2){\circle{2}}
\put(65,2){\circle{2}}
\put(75,2){\circle{2}}
\put(25,2){\circle{2}}
\put(35,2){\circle{2}}
\put(45,2){\circle{2}}
\put(55,2){\circle{2}}
\put(40,-4){\line(0,0){6}}
\qbezier(5,2)(15,17)(25,2)
\qbezier(45,2)(55,17)(65,2)
\qbezier(15,2)(45,26)(75,2)
\qbezier(35,2)(45,17)(55,2)
\end{picture}
\]

{\bf Example 2:}
\begin{eqnarray*}
\begin{picture}(60,10)
\put(0,2){\framebox{\parbox[t]{13.5pt}{$\phantom{\dots}$}}}
\put(5,2){\circle{2}}
\put(15,2){\circle{2}}
\put(25,2){\circle{2}}
\put(30,2){\framebox{\parbox[t]{13.5pt}{$\phantom{\dots}$}}}
\put(35,2){\circle{2}}
\put(45,2){\circle{2}}
\end{picture}
&=&
\begin{picture}(60,10)
\put(0,2){\framebox{\parbox[t]{4.5pt}{$\phantom{\dots}$}}}
\put(0,0.5){\underline{\parbox[b]{50pt}{$\phantom{\dots}$}}}
\put(5,2){\circle{2}}
\put(15,2){\circle{2}}
\qbezier(15,2)(20,12)(25,2)
\put(20,-4){\line(0,0){6}}
\put(25,2){\circle{2}}
\put(30.5,-4){\line(0,0){6}}
\put(30,2){\framebox{\parbox[t]{13.5pt}{$\phantom{\dots}$}}}
\put(35,2){\circle{2}}
\put(45,2){\circle{2}}
\end{picture}
+\quad
\begin{picture}(60,10)
\put(5,2){\circle{2}}
\put(15,2){\circle{2}}
\qbezier(15,2)(20,7)(25,2)
\qbezier(5,2)(20,15)(35,2)
\put(20,-4){\line(0,0){6}}
\put(25,2){\circle{2}}
\put(30,-4){\line(0,0){6}}
\put(35,2){\circle{2}}
\put(45,2){\circle{2}}
\end{picture}
+\quad
\begin{picture}(60,10)
\put(5,2){\circle{2}}
\put(15,2){\circle{2}}
\qbezier(15,2)(25,15)(35,2)
\qbezier(5,2)(15,15)(25,2)
\put(20,-4){\line(0,0){6}}
\put(25,2){\circle{2}}
\put(30,-4){\line(0,0){6}}
\put(35,2){\circle{2}}
\put(45,2){\circle{2}}
\end{picture}
+
\\
&+& 
\begin{picture}(60,30)
\put(0,2){\framebox{\parbox[t]{13.5pt}{$\phantom{\dots}$}}}
\put(0,0.3){\underline{\parbox[b]{50pt}{$\phantom{\dots}$}}}
\put(5,2){\circle{2}}
\put(15,2){\circle{2}}
\put(20,-4){\line(0,0){6}}
\put(25,2){\circle{2}}
\put(30,-4){\line(0,0){6}}
\qbezier(25,2)(30,12)(35,2)
\put(40,2){\framebox{\parbox[t]{4.5pt}{$\phantom{\dots}$}}}
\put(35,2){\circle{2}}
\put(45,2){\circle{2}}
\end{picture}
+\quad
\begin{picture}(60,30)
\put(5,2){\circle{2}}
\put(15,2){\circle{2}}
\qbezier(15,2)(30,15)(45,2)
\put(20,-4){\line(0,0){6}}
\put(25,2){\circle{2}}
\put(30,-4){\line(0,0){6}}
\qbezier(25,2)(30,7)(35,2)
\put(35,2){\circle{2}}
\put(45,2){\circle{2}}
\end{picture}
+\quad
\begin{picture}(60,30)
\put(5,2){\circle{2}}
\put(15,2){\circle{2}}
\qbezier(15,2)(25,15)(35,2)
\put(20,-4){\line(0,0){6}}
\put(25,2){\circle{2}}
\put(30,-4){\line(0,0){6}}
\qbezier(25,2)(35,15)(45,2)
\put(35,2){\circle{2}}
\put(45,2){\circle{2}}
\end{picture}
+\quad
\begin{picture}(60,30)
\put(0,2){\framebox{\parbox[t]{42.5pt}{$\phantom{\dots}$}}}
\put(5,2){\circle{2}}
\put(15,2){\circle{2}}
\put(25,2){\circle{2}}
\put(35,2){\circle{2}}
\put(45,2){\circle{2}}
\end{picture}
\end{eqnarray*}

%%%%%%%%%%%%%%%%%%%%%%%%%%%%%%%%%%%%%%%%%%

\subsection{Formula (\ref{qpWinW})}\label{exqpWinW}

\beqa
\begin{picture}(50,15)
\put(0,2){\framebox{\parbox[t]{42.5pt}{$\phantom{\dots}$}}}
\put(5,2){\circle{2}}
\put(15,2){\circle{2}}
\put(25,2){\circle{2}}
\put(35,2){\circle{2}}
\put(45,2){\circle{2}}
\end{picture}& = &
(\begin{picture}(50,15)
\put(5,2){\circle{2}}
\put(15,2){\circle{2}}
\put(25,2){\circle{2}}
\put(35,2){\circle{2}}
\put(45,2){\circle{2}}
\end{picture})\ +\
(\begin{picture}(50,15)
\put(5,2){\circle{2}}
\put(15,2){\circle{2}}
\put(25,2){\circle{2}}
\put(35,2){\circle{2}}
\put(45,2){\circle{2}}
\qbezier(25,2)(35,17)(45,2)
\end{picture})
\ +\
(\begin{picture}(50,15)
\put(5,2){\circle{2}}
\put(15,2){\circle{2}}
\put(25,2){\circle{2}}
\put(35,2){\circle{2}}
\put(45,2){\circle{2}}
\qbezier(15,2)(25,17)(35,2)
\end{picture})
\ +\
(\begin{picture}(50,15)
\put(5,2){\circle{2}}
\put(15,2){\circle{2}}
\put(25,2){\circle{2}}
\put(35,2){\circle{2}}
\put(45,2){\circle{2}}
\qbezier(5,2)(15,17)(25,2)
\end{picture})
\\&&
\ +\
(\begin{picture}(50,15)
\put(5,2){\circle{2}}
\put(15,2){\circle{2}}
\put(25,2){\circle{2}}
\put(35,2){\circle{2}}
\put(45,2){\circle{2}}
\qbezier(15,2)(30,19)(45,2)
\end{picture})
\ +\
(\begin{picture}(50,15)
\put(5,2){\circle{2}}
\put(15,2){\circle{2}}
\put(25,2){\circle{2}}
\put(35,2){\circle{2}}
\put(45,2){\circle{2}}
\qbezier(5,2)(20,19)(35,2)
\end{picture})
\ +\
(\begin{picture}(50,15)
\put(5,2){\circle{2}}
\put(15,2){\circle{2}}
\put(25,2){\circle{2}}
\put(35,2){\circle{2}}
\put(45,2){\circle{2}}
\qbezier(5,2)(25,19)(45,2)
\end{picture})
\ +\
\begin{picture}(50,15)
\put(5,2){\circle{2}}
\put(15,2){\circle{2}}
\put(25,2){\circle{2}}
\put(35,2){\circle{2}}
\put(45,2){\circle{2}}
\qbezier(5,2)(25,21)(45,2)
\qbezier(15,2)(25,15)(35,2)
\end{picture}
\\&&
\ +\
\begin{picture}(50,15)
\put(5,2){\circle{2}}
\put(15,2){\circle{2}}
\put(25,2){\circle{2}}
\put(35,2){\circle{2}}
\put(45,2){\circle{2}}
\qbezier(5,2)(15,15)(25,2)
\qbezier(15,2)(30,21)(45,2)
\end{picture}
\ +\
\begin{picture}(50,15)
\put(5,2){\circle{2}}
\put(15,2){\circle{2}}
\put(25,2){\circle{2}}
\put(35,2){\circle{2}}
\put(45,2){\circle{2}}
\qbezier(5,2)(20,15)(35,2)
\qbezier(15,2)(30,21)(45,2)
\end{picture}
\ +\
\begin{picture}(50,15)
\put(5,2){\circle{2}}
\put(15,2){\circle{2}}
\put(25,2){\circle{2}}
\put(35,2){\circle{2}}
\put(45,2){\circle{2}}
\qbezier(25,2)(35,15)(45,2)
\qbezier(5,2)(20,21)(35,2)
\end{picture}
\ +\
\begin{picture}(50,15)
\put(5,2){\circle{2}}
\put(15,2){\circle{2}}
\put(25,2){\circle{2}}
\put(35,2){\circle{2}}
\put(45,2){\circle{2}}
\qbezier(15,2)(30,15)(45,2)
\qbezier(5,2)(20,21)(35,2)
\end{picture}
\eeqa
Here, the round brackets denote ordinary Wick ordering (of all
uncontracted fields in an expression).

%%%%%%%%%%%%%%%%%%%%%%%%%%%%%%%%%%%%%%%

\end{appendix}

%%%%%%%%%%%%%%%%%%%%%%%%%%%%%%%%%%%%%%%
%%%%%%%%%%%%%%%%%%%%%%%%%%%%%%%%%%%%%%%

\end{document}